\documentclass[longauth]{aa}

\usepackage{graphicx}
\usepackage{natbib}
\usepackage[english=british]{csquotes}
\usepackage{scalerel}

\usepackage[table]{xcolor}

\usepackage{txfonts}
\usepackage[pdfencoding=auto,psdextra]{hyperref}
\hypersetup{
    colorlinks=true,
    linkcolor=blue,
    filecolor=magenta,      
    urlcolor=blue,
    citecolor=blue
}
\makeatletter
\renewcommand*\aa@pageof{, page \thepage{} of \pageref*{LastPage}}
\makeatother

\usepackage[utf8]{inputenc}

\usepackage[switch, modulo]{lineno}
\usepackage{euclid}
\usepackage{placeins}

\newcommand{\galsim}{\texttt{GalSim}\xspace}
\newcommand{\GEMS}{\texttt{GEMS}\xspace}
\newcommand{\SExtractor}{\texttt{SExtractor}\xspace}
\newcommand{\KSB}{\texttt{KSB}\xspace}

\newcommand\Tstrut{\rule{0pt}{2.6ex}}         
\newcommand\Bstrut{\rule[-0.9ex]{0pt}{0pt}}   

\DeclareSIUnit\mag{mag}
\DeclareSIUnit\sqarcmin{\mathrm{arcmin}^{2}}

\defcitealias{Pujol_2018}{P18}

\begin{document}

\title{\Euclid\/: Improving the efficiency of weak lensing shear bias calibration\thanks{This paper is published on
       behalf of the Euclid Consortium}}
\subtitle{Pixel noise cancellation and the response method on trial}

\newcommand{\orcid}[1]{} 
\author{H.~Jansen$^{1,2}$\thanks{\email{Henning.Jansen@uibk.ac.at}}, M.~Tewes\orcid{0000-0002-1155-8689}$^{1}$, T.~Schrabback\orcid{0000-0002-6987-7834}$^{2,1}$, N.~Aghanim$^{3}$, A.~Amara$^{4}$, S.~Andreon\orcid{0000-0002-2041-8784}$^{5}$, N.~Auricchio\orcid{0000-0003-4444-8651}$^{6}$, M.~Baldi\orcid{0000-0003-4145-1943}$^{7,6,8}$, E.~Branchini\orcid{0000-0002-0808-6908}$^{9,10}$, M.~Brescia\orcid{0000-0001-9506-5680}$^{11,12}$
% ...
}

\institute{$^{1}$ Universit\"at Bonn, Argelander-Institut f\"ur Astronomie, Auf dem H\"ugel 71, 53121 Bonn, Germany\\
$^{2}$ Universit\"at Innsbruck, Institut f\"ur Astro- und Teilchenphysik, Technikerstr. 25/8, 6020 Innsbruck, Austria\\
$^{3}$ Universit\'e Paris-Saclay, CNRS, Institut d'astrophysique spatiale, 91405, Orsay, France\\
$^{4}$ Institute of Cosmology and Gravitation, University of Portsmouth, Portsmouth PO1 3FX, UK\\
$^{5}$ INAF-Osservatorio Astronomico di Brera, Via Brera 28, 20122 Milano, Italy\\
$^{6}$ INAF-Osservatorio di Astrofisica e Scienza dello Spazio di Bologna, Via Piero Gobetti 93/3, 40129 Bologna, Italy\\
$^{7}$ Dipartimento di Fisica e Astronomia, Universit\'a di Bologna, Via Gobetti 93/2, 40129 Bologna, Italy\\
$^{8}$ INFN-Sezione di Bologna, Viale Berti Pichat 6/2, 40127 Bologna, Italy\\
$^{9}$ Dipartimento di Fisica, Universit\'a di Genova, Via Dodecaneso 33, 16146, Genova, Italy\\
$^{10}$ INFN-Sezione di Genova, Via Dodecaneso 33, 16146, Genova, Italy\\
$^{11}$ Department of Physics "E. Pancini", University Federico II, Via Cinthia 6, 80126, Napoli, Italy\\
$^{12}$ INAF-Osservatorio Astronomico di Capodimonte, Via Moiariello 16, 80131 Napoli, Italy\\
$^{13}$ Instituto de Astrof\'isica e Ci\^encias do Espa\c{c}o, Universidade do Porto, CAUP, Rua das Estrelas, PT4150-762 Porto, Portugal\\
$^{14}$ Dipartimento di Fisica, Universit\'a degli Studi di Torino, Via P. Giuria 1, 10125 Torino, Italy\\
$^{15}$ INFN-Sezione di Torino, Via P. Giuria 1, 10125 Torino, Italy\\
$^{16}$ INAF-Osservatorio Astrofisico di Torino, Via Osservatorio 20, 10025 Pino Torinese (TO), Italy\\
$^{17}$ INAF-IASF Milano, Via Alfonso Corti 12, 20133 Milano, Italy\\
$^{18}$ INAF-Osservatorio Astronomico di Roma, Via Frascati 33, 00078 Monteporzio Catone, Italy\\
$^{19}$ INFN-Sezione di Roma, Piazzale Aldo Moro, 2 - c/o Dipartimento di Fisica, Edificio G. Marconi, 00185 Roma, Italy\\
$^{20}$ Institut de F\'{i}sica d'Altes Energies (IFAE), The Barcelona Institute of Science and Technology, Campus UAB, 08193 Bellaterra (Barcelona), Spain\\
$^{21}$ Port d'Informaci\'{o} Cient\'{i}fica, Campus UAB, C. Albareda s/n, 08193 Bellaterra (Barcelona), Spain\\
$^{22}$ Institute for Theoretical Particle Physics and Cosmology (TTK), RWTH Aachen University, 52056 Aachen, Germany\\
$^{23}$ INFN section of Naples, Via Cinthia 6, 80126, Napoli, Italy\\
$^{24}$ Dipartimento di Fisica e Astronomia "Augusto Righi" - Alma Mater Studiorum Universit\'a di Bologna, Viale Berti Pichat 6/2, 40127 Bologna, Italy\\
$^{25}$ Institute for Astronomy, University of Edinburgh, Royal Observatory, Blackford Hill, Edinburgh EH9 3HJ, UK\\
$^{26}$ European Space Agency/ESRIN, Largo Galileo Galilei 1, 00044 Frascati, Roma, Italy\\
$^{27}$ ESAC/ESA, Camino Bajo del Castillo, s/n., Urb. Villafranca del Castillo, 28692 Villanueva de la Ca\~nada, Madrid, Spain\\
$^{28}$ University of Lyon, Univ Claude Bernard Lyon 1, CNRS/IN2P3, IP2I Lyon, UMR 5822, 69622 Villeurbanne, France\\
$^{29}$ Institute of Physics, Laboratory of Astrophysics, Ecole Polytechnique F\'ed\'erale de Lausanne (EPFL), Observatoire de Sauverny, 1290 Versoix, Switzerland\\
$^{30}$ UCB Lyon 1, CNRS/IN2P3, IUF, IP2I Lyon, 4 rue Enrico Fermi, 69622 Villeurbanne, France\\
$^{31}$ Departamento de F\'isica, Faculdade de Ci\^encias, Universidade de Lisboa, Edif\'icio C8, Campo Grande, PT1749-016 Lisboa, Portugal\\
$^{32}$ Instituto de Astrof\'isica e Ci\^encias do Espa\c{c}o, Faculdade de Ci\^encias, Universidade de Lisboa, Campo Grande, 1749-016 Lisboa, Portugal\\
$^{33}$ Department of Astronomy, University of Geneva, ch. d'Ecogia 16, 1290 Versoix, Switzerland\\
$^{34}$ INAF-Istituto di Astrofisica e Planetologia Spaziali, via del Fosso del Cavaliere, 100, 00100 Roma, Italy\\
$^{35}$ Universit\'e Paris-Saclay, Universit\'e Paris Cit\'e, CEA, CNRS, AIM, 91191, Gif-sur-Yvette, France\\
$^{36}$ INAF-Osservatorio Astronomico di Trieste, Via G. B. Tiepolo 11, 34143 Trieste, Italy\\
$^{37}$ INAF-Osservatorio Astronomico di Padova, Via dell'Osservatorio 5, 35122 Padova, Italy\\
$^{38}$ Max Planck Institute for Extraterrestrial Physics, Giessenbachstr. 1, 85748 Garching, Germany\\
$^{39}$ University Observatory, Faculty of Physics, Ludwig-Maximilians-Universit{\"a}t, Scheinerstr. 1, 81679 Munich, Germany\\
$^{40}$ Institute of Theoretical Astrophysics, University of Oslo, P.O. Box 1029 Blindern, 0315 Oslo, Norway\\
$^{41}$ Leiden Observatory, Leiden University, Niels Bohrweg 2, 2333 CA Leiden, The Netherlands\\
$^{42}$ Jet Propulsion Laboratory, California Institute of Technology, 4800 Oak Grove Drive, Pasadena, CA, 91109, USA\\
$^{43}$ von Hoerner \& Sulger GmbH, Schlo{\ss}Platz 8, 68723 Schwetzingen, Germany\\
$^{44}$ Technical University of Denmark, Elektrovej 327, 2800 Kgs. Lyngby, Denmark\\
$^{45}$ Cosmic Dawn Center (DAWN), Denmark\\
$^{46}$ Institut d'Astrophysique de Paris, UMR 7095, CNRS, and Sorbonne Universit\'e, 98 bis boulevard Arago, 75014 Paris, France\\
$^{47}$ Max-Planck-Institut f\"ur Astronomie, K\"onigstuhl 17, 69117 Heidelberg, Germany\\
$^{48}$ Department of Physics and Astronomy, University College London, Gower Street, London WC1E 6BT, UK\\
$^{49}$ Aix-Marseille Universit\'e, CNRS/IN2P3, CPPM, Marseille, France\\
$^{50}$ Universit\'e Paris-Saclay, Universit\'e Paris Cit\'e, CEA, CNRS, Astrophysique, Instrumentation et Mod\'elisation Paris-Saclay, 91191 Gif-sur-Yvette, France\\
$^{51}$ Mullard Space Science Laboratory, University College London, Holmbury St Mary, Dorking, Surrey RH5 6NT, UK\\
$^{52}$ Department of Physics, P.O. Box 64, 00014 University of Helsinki, Finland\\
$^{53}$ Helsinki Institute of Physics, Gustaf H{\"a}llstr{\"o}min katu 2, University of Helsinki, Helsinki, Finland\\
$^{54}$ NOVA optical infrared instrumentation group at ASTRON, Oude Hoogeveensedijk 4, 7991PD, Dwingeloo, The Netherlands\\
$^{55}$ Aix-Marseille Universit\'e, CNRS, CNES, LAM, Marseille, France\\
$^{56}$ Dipartimento di Fisica e Astronomia "Augusto Righi" - Alma Mater Studiorum Universit\'a di Bologna, via Piero Gobetti 93/2, 40129 Bologna, Italy\\
$^{57}$ Department of Physics, Institute for Computational Cosmology, Durham University, South Road, DH1 3LE, UK\\
$^{58}$ Universit\'e Paris Cit\'e, CNRS, Astroparticule et Cosmologie, 75013 Paris, France\\
$^{59}$ University of Applied Sciences and Arts of Northwestern Switzerland, School of Engineering, 5210 Windisch, Switzerland\\
$^{60}$ Institut d'Astrophysique de Paris, 98bis Boulevard Arago, 75014, Paris, France\\
$^{61}$ CEA Saclay, DFR/IRFU, Service d'Astrophysique, Bat. 709, 91191 Gif-sur-Yvette, France\\
$^{62}$ Department of Physics, Oxford University, Keble Road, Oxford OX1 3RH, UK\\
$^{63}$ European Space Agency/ESTEC, Keplerlaan 1, 2201 AZ Noordwijk, The Netherlands\\
$^{64}$ Department of Physics and Astronomy, University of Aarhus, Ny Munkegade 120, DK-8000 Aarhus C, Denmark\\
$^{65}$ Space Science Data Center, Italian Space Agency, via del Politecnico snc, 00133 Roma, Italy\\
$^{66}$ Centre National d'Etudes Spatiales -- Centre spatial de Toulouse, 18 avenue Edouard Belin, 31401 Toulouse Cedex 9, France\\
$^{67}$ Dipartimento di Fisica e Astronomia "G. Galilei", Universit\'a di Padova, Via Marzolo 8, 35131 Padova, Italy\\
$^{68}$ INFN-Padova, Via Marzolo 8, 35131 Padova, Italy\\
$^{69}$ Universit\"ats-Sternwarte M\"unchen, Fakult\"at f\"ur Physik, Ludwig-Maximilians-Universit\"at M\"unchen, Scheinerstrasse 1, 81679 M\"unchen, Germany\\
$^{70}$ Departamento de F\'isica, FCFM, Universidad de Chile, Blanco Encalada 2008, Santiago, Chile\\
$^{71}$ Institut d'Estudis Espacials de Catalunya (IEEC), Carrer Gran Capit\'a 2-4, 08034 Barcelona, Spain\\
$^{72}$ Institute of Space Sciences (ICE, CSIC), Campus UAB, Carrer de Can Magrans, s/n, 08193 Barcelona, Spain\\
$^{73}$ Satlantis, University Science Park, Sede Bld 48940, Leioa-Bilbao, Spain\\
$^{74}$ Centre for Electronic Imaging, Open University, Walton Hall, Milton Keynes, MK7~6AA, UK\\
$^{75}$ Centro de Investigaciones Energ\'eticas, Medioambientales y Tecnol\'ogicas (CIEMAT), Avenida Complutense 40, 28040 Madrid, Spain\\
$^{76}$ Instituto de Astrof\'isica e Ci\^encias do Espa\c{c}o, Faculdade de Ci\^encias, Universidade de Lisboa, Tapada da Ajuda, 1349-018 Lisboa, Portugal\\
$^{77}$ Universidad Polit\'ecnica de Cartagena, Departamento de Electr\'onica y Tecnolog\'ia de Computadoras,  Plaza del Hospital 1, 30202 Cartagena, Spain\\
$^{78}$ Institut de Recherche en Astrophysique et Plan\'etologie (IRAP), Universit\'e de Toulouse, CNRS, UPS, CNES, 14 Av. Edouard Belin, 31400 Toulouse, France\\
$^{79}$ Kapteyn Astronomical Institute, University of Groningen, PO Box 800, 9700 AV Groningen, The Netherlands\\
$^{80}$ INFN-Bologna, Via Irnerio 46, 40126 Bologna, Italy\\
$^{81}$ Infrared Processing and Analysis Center, California Institute of Technology, Pasadena, CA 91125, USA\\
$^{82}$ Instituto de Astrof\'isica de Canarias, Calle V\'ia L\'actea s/n, 38204, San Crist\'obal de La Laguna, Tenerife, Spain\\
$^{83}$ Junia, EPA department, 41 Bd Vauban, 59800 Lille, France}

\abstract{To obtain an accurate cosmological inference from upcoming weak lensing surveys such as the one conducted by \Euclid, the shear measurement requires calibration using galaxy image simulations. As it typically requires millions of simulated galaxy images and consequently a substantial computational effort, seeking methods to speed the calibration up is valuable. We study the efficiency of different noise cancellation methods that aim at reducing the simulation volume required to reach a given precision in the shear measurement. The more efficient a method is, the faster we can estimate the relevant biases up to a required precision level. Explicitly, we compared fit methods with different noise cancellations and a method based on responses. We used \galsim to simulate galaxies both on a grid and at random positions in larger scenes. Placing the galaxies at random positions requires their detection, which we performed with \SExtractor. On the grid, we neglected the detection step and, therefore, the potential detection bias arising from it. The shear of the simulated images was measured with the fast moment-based method \KSB, for which we note deviations from purely linear shear measurement biases. For the estimation of uncertainties, we used bootstrapping as an empirical method. We extended the response-based approach to work on a wider range of shears and provide accurate estimates of selection biases. We find that each method we studied on top of shape noise cancellation can further increase the efficiency of calibration simulations. The improvement depends on the considered shear amplitude range and the type of simulations (grid-based or random positions). The response method on a grid for small shears provides the biggest improvement. Here the runtime for the estimation of multiplicative biases can be lowered by a factor of $145$ compared to the benchmark simulations without any cancellation. In the more realistic case of randomly positioned galaxies, we still find an improvement factor of $70$ for small shears using the response method. Alternatively, the runtime can be lowered by a factor of 7 already using pixel noise cancellation on top of shape noise cancellation. Furthermore, we demonstrate that the efficiency of shape noise cancellation can be enhanced in the presence of blending if entire scenes are rotated instead of individual galaxies. }
  
\keywords{Gravitational lensing: weak --
            Methods: data analysis
           }

\titlerunning{\Euclid: Improving the efficiency of shear bias calibration.}
\authorrunning{H. Jansen et al.}

\maketitle
   
\section{Introduction}
According to the dark energy task force, weak lensing is one of the four most promising methods to constrain the equation of state of dark energy \citep[see][]{WL_taskforce}. In particular, weak lensing has the potential to be the most powerful method alongside these four suggested methods if systematic biases can be controlled with sufficient accuracy. For \Euclid \citep{2011arXiv1110.3193L}, the Nancy Grace Roman Space Telescope \citep{2015arXiv150303757S}, and other weak lensing surveys like the Kilo-Degree Survey (KiDS) \citep{2013Msngr.154...44D}, the Dark Energy Survey \citep{DES}, and the Hyper Suprime-cam (HSC) survey \citep{HSC}, it is therefore inevitable that calibration or validation of weak lensing shear measurement algorithms is required. For this calibration, galaxy image simulations are used, that are as close as possible to the real images produced by those instruments \citep[see][]{2015MNRAS.449..685H, 2017MNRAS.468.3295H}. Recent image simulations for the aforementioned surveys \citep[see][]{SKiLLS, 2022MNRAS.509.3371M, 2019A&A...624A..92K, 2018MNRAS.481.3170M} therefore carefully adjust their input parameters such that they match actual observations. These image simulations require $10^7$--$10^{10}$ simulated galaxy images to achieve the desired uncertainty levels on the biases \citep[see][]{RefId0}. Since that is computationally expensive, efforts are put into reducing the required simulated galaxies. The main approach used in several studies \citep[see][]{shape_noise_cancel, Mandelbaum, Pujol_2018} is shape noise cancellation. This cancellation aims to remove the noise introduced by the intrinsic shape distribution of the input galaxy catalog. In most cases, this is done by considering an additional image with the same galaxy rotated by $90$ degrees, but it is also possible to perform this cancellation in a full ring \citep[see][]{Nakajima_2007}. This kind of cancellation does not work perfectly due to additional pixel shot noise on the image and the pixelation itself. In particular, when the galaxies are not isolated, blending also degrades the efficiency of this cancellation, as shown in \citet{2021A&A...646A.124H}. Still it has proven to be useful in the simulations. In addition the pixel noise realisation on an image might make the galaxy look slightly rounder or more elliptical for the considered shape measurement method. \citet{2012MNRAS.424.2757M} show that pixel noise gives rise to biases for the shear measurement. This is why already \citet{RefId0} used another cancellation, which utilised an inverted pixel noise realisation. Having two images, where one has an inverted noise realisation, can potentially help to cancel the previously discussed effect. The authors referred to this cancellation as background noise cancellation, while we use the name pixel noise cancellation in this paper. We studied this method in more detail and worked out the benefits of this method in different scenarios. In particular, we studied the effects of cancelling also the shot noise from the galaxy itself and not just a white Gaussian noise field. These mentioned cancellations are all applicable when determining shear bias parameters from the commonly used regression of measured shear as a function of true input shear. 

Another approach is suggested by \citet[hereafter P18]{Pujol_2018} requiring a different setup of the simulations, but potentially for a large reward in terms of runtime improvement. In this method, the biases were determined from the individual shear responses of galaxies. To obtain these responses, we simulated each galaxy with two slightly differing shears. The individual responses are very noisy, but averaging over many of those can yield reliable estimates of the systematic biases. As this method, by definition, uses individual galaxies, it also makes it very easy to study the effects of specific galaxy properties like the Sérsic index or half-light radius on the biases. The authors' original approach does not account for selection effects, so we expanded their formalism based on ideas suggested in their paper to also account for selection effects and larger shear intervals. Also here we accounted for the complete pixel noise including shot noise, while \citetalias{Pujol_2018} only used a white Gaussian noise field. 

This paper scrutinises all these methods in two fundamentally different scenarios, with simulations roughly mimicking \Euclid observing conditions. As a first step, we simulated isolated galaxies down to $\SI{24.5}{\mag}$ on a grid. Most of the galaxies in the input catalog are also in the output catalog as no detection step was needed and most of the galaxies have a signal-to-noise ratio of more than 10, which was chosen as a selection criterion. As a second step, we placed galaxies at random positions embedded in a larger scene. We included galaxies down to $\SI{26.5}{\mag}$ in these scenes such that there is also blending by undetected sources. 

We summarise basic weak lensing formalism in Sect.~\ref{sec:2}. In Sect.~\ref{sec:3} we describe the methods studied in this paper to reduce noise and speed up the simulations. Section~\ref{sec:4} presents the setup of the simulations in more detail. Section~\ref{sec:5} describes the uncertainty on the bias parameters, which is crucial for the subsequent efficiency comparison. In Sect.~\ref{sec:6} we present our results comparing the uncertainties of the different methods with their respective runtimes. Finally in Sect.~\ref{sec:7} we summarise our results and briefly discuss the implications for future calibration simulations.

\section{Weak lensing formalism}\label{sec:2}
    \subsection{Definition and measurement of shear}
        In weak lensing, we can linearise the lens mapping if the angular scale of the lensed image is smaller than the scale on which the tidal field varies \citep[see][]{Schneider_2006}. This lens mapping is then given by the Jacobian $\mathit{A}$ defined as
        \begin{equation}\label{eq:Jacobi}
            \mathit{A} = (1-\kappa)\begin{pmatrix} 1-g_1 & -g_2 \\ -g_2 & 1+g_1 \end{pmatrix}\,.
        \end{equation}
        Here $\kappa$ is the convergence describing the change of the apparent size and $g_i$ denotes the component $i$ of the reduced shear $g$, which is defined as 
        \begin{equation}\label{eq:reduced_shear_def}
            g = \frac{\gamma}{1-\kappa} \,,
        \end{equation}
        with the true shear $\gamma$. In Eq.~\eqref{eq:reduced_shear_def} the standard notation as a complex number $g = g_1 + \mathrm{i}\,g_2$ is used. We can only measure the reduced shear directly. Still in the case of weak lensing the convergence is typically small such that the reduced shear is about the size of the true shear. Therefore we refer to the reduced shear as just the shear for the rest of the paper. The shear and ellipticity in general are invariant under rotations of $\pi$ radians. Therefore we can characterise these quantities as spin-2 \citep[see][]{2005PhRvD..72b3516C}. Assuming that the intrinsic ellipticities have no preferential alignment, the observed ellipticities $\epsilon_\mathrm{obs}$ give an unbiased estimate of the shear:
        \begin{equation}\label{eq:shear_estimate}
            g = \langle \epsilon_{\mathrm{obs}}\rangle\,.
        \end{equation}
        This equation only holds for specific definitions of ellipticity. For a light distribution with elliptical isophotes, such an ellipticity definition is
        \begin{equation}
            |\epsilon| = \frac{1-r}{1+r}\,,
        \end{equation}
        where $r$ denotes the axis ratio $b / a$ of the ellipse with $b$ being the semi-minor axis and $a$ being the semi-major axis \citep[see][]{Schneider_2006}. Still an ellipticity definition, for which Eq.~\eqref{eq:shear_estimate} holds, can also be found for non-elliptical galaxies. We performed the ellipticity measurement with the \texttt{HSM} module from the \galsim library \citep[see][]{galsim}. The HSM module uses algorithms from \citet{2003MNRAS.343..459H} probed on real data from \citet{2005MNRAS.361.1287M}. It also has a specific version of the \KSB \citep{KSB, 1997ApJ...475...20L, 1998ApJ...504..636H} shape measurement method implemented, which uses weighted brightness moments to estimate the ellipticity and correct for the PSF. 
        
        For simplicity we only studied the behaviour for the first shear component $g_1$ and set $g_2=0$. We find no significant correlation between biases for the two shear components in our analysis such that an individual study of each component is valid.

    \subsection{Determination of shear bias}\label{sec:det_bias}
        The biases of such a shear estimator can then be studied with different methods. To first order a linear bias model as described in \citet{2006MNRAS.368.1323H} is used in this analysis. Such a linear bias model has been used for the majority of weak lensing studies to date starting from \citet{2005PhRvD..72d3503G} and \citet{2006MNRAS.366..101H}. The difference between a shear estimator $\hat{g}$ and the respective true shear $g^{\mathrm{true}}$ following this model is given by
        \begin{equation}\label{eq:bias_definition}
            \hat{g_i} - g^{\mathrm{true}}_{i} = \mu_i \, g_i^{\mathrm{true}} + c_i + \text{noise}\,,
        \end{equation}
        where $i=\{1,2\}$. The multiplicative part $\mu$ is referred to in the following as the multiplicative bias parameter and the additive part $c$ as the additive bias parameter. In general if the biases for both shear components are independent, Eq.~\eqref{eq:bias_definition} does not conserve spin \citep[see][]{kitching}. These parameters can be determined by simulating galaxies with different but known true shears and then fitting a straight line to the shear estimation residuals against the true shear. In order for this bias estimate to be precise, we need to simulate a large sample of galaxies per constant input shear to average out the intrinsic ellipticities. 
        
        \citet{Sheldon_2017} suggest a new formalism for the shear bias determination based on shear responses. Assuming we have an estimate $e_i$ for component $i$ of the complex ellipticity $e$, this estimate can be expanded in a Taylor series as 
        \begin{equation}
            e_i \approx \left. e_i \right|_{\gamma=0} + \sum_{j=1}^{2} \left. \frac{\partial e_i}{\partial \gamma_j} \right|_{\gamma=0} \gamma_j + \dots\,.
        \end{equation}
        The authors define the shear response as
        \begin{equation}
            R_{ij} = \left.\frac{\partial e_i}{\partial \gamma_j}\right|_{\gamma=0}\,.
        \end{equation}
        In a large enough sample, the average of the intrinsic ellipticities given by the first term in the Taylor expansion vanishes such that 
        \begin{equation}
            \langle e_i \rangle \approx \left\langle \sum_{j=1}^{2} R_{ij}\, \gamma_j \right\rangle\,.
        \end{equation}
        The shear response matrix $\boldsymbol{R}$ is used by \citetalias{Pujol_2018} to estimate the $\mu$ and $c$ biases. In the following, we refer to the response method as RM. We sometimes append a number to this abbreviation, which stands for the maximum amplitude of the shear used in that specific case. The response can be measured for individual galaxies and is a $2\times 2$ matrix 
        \begin{equation}
            \boldsymbol{R} = \begin{pmatrix}
                            R_{11} & R_{12} \\
                            R_{21} & R_{22}\\ 
                            \end{pmatrix} \,,
        \end{equation}
        where the diagonal terms directly translate to the multiplicative bias via 
        \begin{equation}
            1 + \mu_{i} = R_{ii}\,,
        \end{equation} 
        while the cross-terms express the correlation between the two ellipticity components. Following \citet{2023arXiv230214656K} these off-diagonal terms represent a mixture of spin-0 and spin-4 terms. As we focus on $g_1$ in this paper, only $R_{11}$ is relevant for us and the indices are dismissed in the following. The determination of this response requires simulated images of the same galaxy with a finite shear difference, as described later in Sect.~\ref{sec:3_rm}. 
        
        While the additive bias $c$ is just another fit parameter for the linear regression technique, it is not as easy to determine from the response method. Following \citet[eq. 3]{Pujol_2018} we can determine individual additive biases via 
        \begin{equation}
            a_i = e_i^{\mathrm{obs}} - R_{ii} \, g_i - e_i^\mathrm{I}\,,
        \end{equation}
        where $e_i^\mathrm{I}$ denotes the intrinsic ellipticity. The $c$-bias can then be determined as the average of the $a_i$ for many simulated galaxy images. This approach is biased since the response $R_{ii}$ needs to be calculable. Thus, we do not want to use this approach in our study. Still, we can estimate the $c$ bias from $c_i= \langle e_i^\mathrm{obs}\rangle$ if the input ellipticity vanishes on average. In Sect.~\ref{sec:4_setup}, we describe how this is ensured in more detail.

\section{Methods to reduce the impact of noise}\label{sec:3}
    The measurement of shear is dominated by noise. The central part of this noise is the shape noise due to the unknown intrinsic ellipticity of each galaxy. Furthermore an image taken by a CCD camera contains pixel noise with at least three contributions: read-out noise (here assumed to be Gaussian), Poisson shot noise from the sky background, and Poisson noise coming from the galaxies' flux. In the case of \Euclid, the dominant part of the pixel noise is the contribution from the sky background, which is why past simulations have often only considered this stationary part. In this paper, however, we systematically include all the noise sources mentioned above in our model as they affect the shear bias estimation. 
    
    Due to its dominant role, it makes sense to think of methods to mitigate noise and increase the efficiency of simulations. In the following, we list possible methods studied in this paper. 
    \subsection{Fit method}
\begin{figure}
            \resizebox{0.49\hsize}{!}{\includegraphics{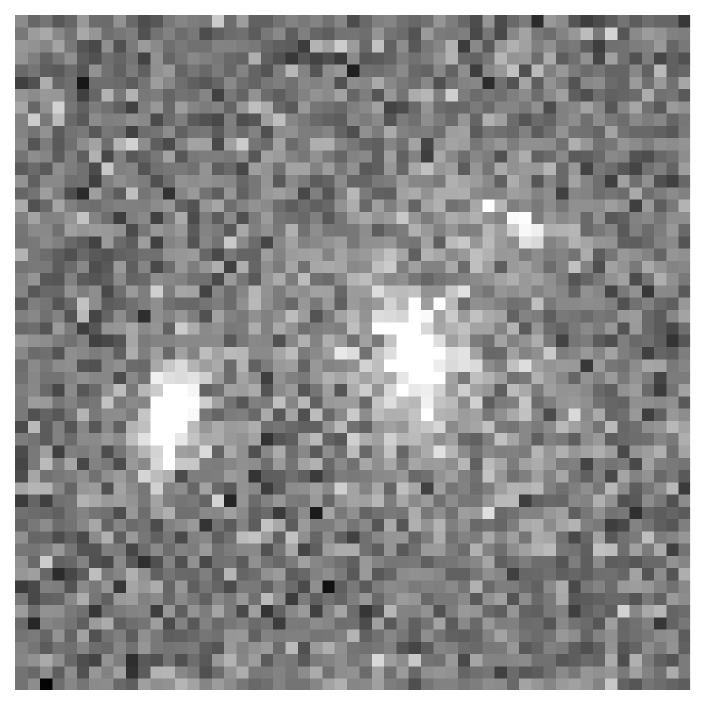}}
            \resizebox{0.49\hsize}{!}{\includegraphics{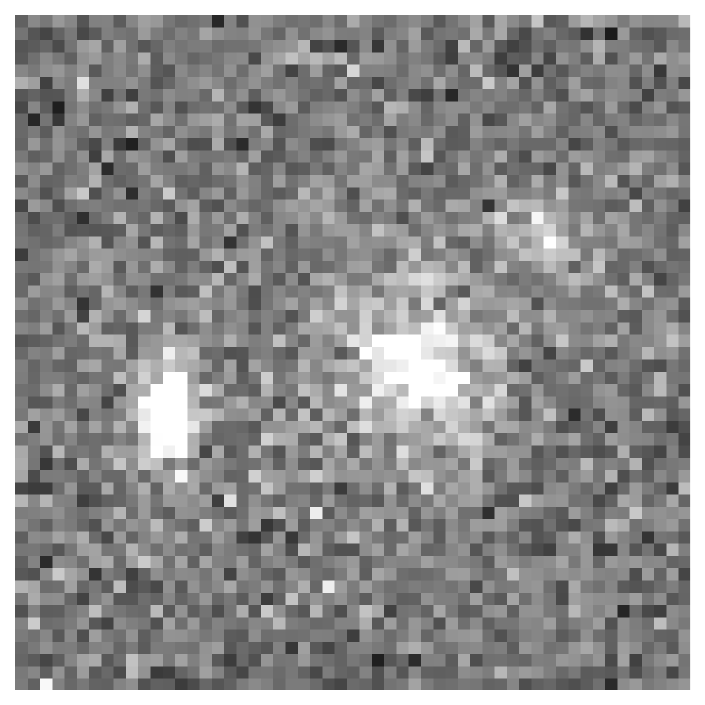}}
            \caption{Effect of pixel noise on three simulated faint galaxies. While the pixel noise realisation is added in the left panel, it is inverted in the right panel. The pixel noise realisation clearly affects the apparent shape (especially the shape of the faint galaxy on the top right).} 
            \label{fig:pixel_noise_example}
        \end{figure}

        The fit method determines shear biases by fitting a model to the estimated shears as discussed in Sect.~\ref{sec:det_bias}. Here the idea is to reduce the uncertainty on the estimated shears directly by cancelling the most important noise components within groups of simulated images. One is the shape noise induced by the intrinsic distribution of ellipticities. Typically the intrinsic ellipticity is an order of magnitude larger than the shear signal. Thus individual shear estimates are dominated by this intrinsic shape. The second essential component is pixel noise. One realisation of this noise might make the galaxy look a bit rounder (or more elliptical) than it actually is. In Fig.~\ref{fig:pixel_noise_example}, the effect of pixel noise is shown for one galaxy. The cancellation of these noise components is discussed in the following. 
        
        \subsubsection{Shape noise cancellation}
            One popular method to reduce the impact of intrinsic shapes is shape noise cancellation \citep[see][]{shape_noise_cancel, Mandelbaum, Pujol_2018}. This method uses a second image of the same galaxy, which is rotated by $90$ degrees with respect to the original image. Without any noise and selection effects, this would cancel out the intrinsic shapes and the average measured ellipticity would be the shear. In reality this rotation does not cancel out shape noise perfectly, but it still improves the performance significantly. In a related effort, one can also use more than two galaxies for the rotation. This method was suggested by \citet{Nakajima_2007} as the \enquote{ring test}. Our study focuses on the simplest case with just one $90$ degree rotated pair. \citet{2017MNRAS.467.1627F} find that for KiDS using four rotated versions improves the cancellation, but \citet{RefId0} show that using more than two versions does not yield further improvements under \Euclid conditions, which we adapted in this work.  
        
        \subsubsection{Pixel noise cancellation}\label{sec:pixel_noise_cancellation}
            On top of shape noise cancellation, we want to evaluate what we refer to as pixel noise cancellation. The idea is to build a second version of each simulated image using an inverted noise field. Using this cancellation, a noise field, which makes the galaxy look rounder in one version, possibly has the opposite effect in the second version. That way the impact of pixel noise on the shear measurement might be reduced and the efficiency might be increased. With the requirements of previous surveys, it would have been sufficient just to subtract the exact same noise, which was added before\footnote{Extracting the noise can be done by subtracting the image with noise from the image without noise using \galsim \citep[see][]{galsim}.}. This procedure inherently assumes that the noise is exactly symmetric. Usually this can be assumed since the Poisson distribution approaches a Gaussian distribution for high counts. To make sure that we do not introduce any additional bias at the level of the very tight requirements for \Euclid, we did not just subtract the noise realisation but inverted the noise properly. For the Gaussian read-out noise, this can indeed be done by just extracting the noise realisation and subtracting it instead of adding it. For the Poisson noise, we wanted the inverted realisation to follow the same Poisson distribution. Since the \galsim implementation of Poisson noise shows a chaotic behaviour when using the same seed with slightly different means, we implemented the Poisson noise ourselves using the inverse transform method. For the noise in one pixel, we generated a random number $U$ between zero and one and then summed up the cumulative distribution function until it exceeded the drawn random number. The inverse noise realisation can then be found by doing the same for $1-U$. Thus, a positive noise realisation in the $90$-th percentile of the Poisson distribution has a negative counterpart in the $10$-th percentile of the same distribution. Given that the Poisson distribution is not exactly symmetric, these two drawn realisations are not symmetric around the mean. Therefore our implementation does in fact not exactly \enquote{invert} the sign of the noise. 
            
            This method requires that the galaxy is at exactly the same position in the image. For shape noise cancellation, we are free to change for example sub-pixel shifts between the pairs. However it made the cancellation less effective, as we drew new noise realisations anyway for the rotated stamp. But for pixel noise cancellation, the counts in a pixel, which generated the noise realisation in the first place, need to be still associated with their noise realisation in the second version. Pixel noise cancellation is relatively cheap, as no additional convolution is needed to build a different version of the galaxy. 
    
    \subsection{Response method}\label{sec:3_rm}
         The shear response is determined by the difference of the observed ellipticities for the same galaxy divided by the finite shear difference between them. Thus this response can be estimated by choosing a small interval around zero $[-\Delta g, \Delta g]$ and simulating one image for each margin of the interval. That way two images are built and the response can be estimated as
        \begin{equation}
            R \approx \frac{e_{\mathrm{obs}}^+ - e_\mathrm{{obs}}^-}{2\Delta g}\,,
        \end{equation}
        where $e_{\mathrm{obs}}$ denotes the measured ellipticity from \KSB \citepalias[see][eq. 4]{Pujol_2018}. The $+$ or $-$ in superscript indicates that the shear has been increased or decreased respectively. Averaging all individual responses can then give an estimate for the multiplicative bias, but this would not include the selection bias, as a detection or selection yields incomplete pairs for which the response can not be computed. To include the selection bias, \citetalias{Pujol_2018} suggest building the response as
        \begin{equation}\label{eq:pujol_33}
            \langle R \rangle \approx \frac{\langle e_{\mathrm{obs}}^+\rangle - \langle e_{\mathrm{obs}}^-\rangle }{2\Delta g}\,.
        \end{equation}
        The $\langle e_{\mathrm{obs}} \rangle$ enclose all measurements even if their respective partner was not measured. This estimate inherently accounts for the unavoidable selection bias and needs to be considered for the comparison with other methods. 
        
        In their original paper, \citetalias{Pujol_2018} used the same Gaussian noise realisation for both images, which is crucial to stabilising the method.
        Indeed we find that drawing an independent realisation of the CCD noise for each image destabilises the method so that it is not usable anymore. To keep the noise as similar as possible within a pair, we used the same seed for the noise generation of the images belonging to each other. Since the same seed with slightly different means does not produce the same noise pattern using the \galsim Poisson noise generator, we again used the inverse transform method as described in Sect.~\ref{sec:pixel_noise_cancellation} to generate the noise. The mean of the distribution might then change due to the different shear, which makes some pixel gain flux and others lose flux. However, the noise realisation is still generated from the same percentile of the Poisson distribution since the seed fixes the random number used to generate the realisation. 
        
        As discussed later in this paper, the size of the used interval starts to play a role for larger shears of several percent. Thus the original response method with a $\Delta g=0.02$ can not be used when comparing to fit methods in an interval $[-0.1, 0.1]$, as the recovered biases would differ significantly. We therefore extended the method using 11 differently sheared images evenly spaced between $-0.1$ and $0.1$. In this way, we can estimate ten responses covering the same interval as the other methods while keeping the $\Delta g$ between two images small enough. A $\Delta g$ which is too large distorts the results as the selection effects are more important for larger shear differences. \citetalias{Pujol_2018} state the upper limit for $\Delta g$ in their analysis to be $0.05$.

\section{Simulated data}\label{sec:4}
    In this paper, we study the behaviour of different shear estimation methods in two scenarios. We began with the easiest case with galaxies on a grid without any blends. In the second scenario, we then placed galaxies at random positions on $4000\times 4000$ pixel large scenes. Thus we could also study how blending affects the different methods. This treatment without a grid also includes the usage of \SExtractor in the detection step as introduced by \citet{Sextractor}. The highly simplified simulations described in the following are solely used for this study and should not be seen as representative of the images expected from the \Euclid VIS instrument.
    
    \subsection{Setup of the simulations}\label{sec:4_setup}
        We constructed the simulations in this paper with \galsim, which was first introduced in \citet{galsim}. In order to parallelise the code, we used the Python library Ray, which was first introduced by \citet{ray}. The galaxies were drawn from the \GEMS catalog \citep{GEMS}. This survey was conducted using the \textit{Hubble} Space Telescope (HST). We used a comparable selection\footnote{The single further restriction in this work is the lower limit on half-light radius. We required the half-light radius to be at least one \Euclid pixel large, which corresponds to a little more than three \GEMS pixels.} as \citet{Tewes_2019} and refer the reader to their paper for the details. Using their selection criteria, we ended up with $9026$ galaxies for the grid-based simulations with the faintest galaxies having $\SI{24.5}{\mag}$ and $36438$ galaxies for the large-scene simulations with magnitudes as faint as $\SI{26.5}{\mag}$. Only the magnitude, the Sérsic index, and the half-light radius were taken from the \GEMS catalog. The absolute values of the intrinsic ellipticities were drawn from a truncated Rayleigh distribution with $\sigma_\epsilon = 0.25$, where the ellipticity definition $|\epsilon| = (1-r) / (1+r)$ was used. We truncated the distribution at $|\epsilon|=0.7$ to avoid convolution problems with highly elliptical galaxies. Additionally an orientation angle was drawn from a uniform distribution. In this way, we avoided any correlation between the half-light radius and ellipticity used as input to our simulations. The galaxies were drawn as single Sérsic profiles and made elliptical using the area-preserving shearing of \galsim. We note that if accurate absolute biases shall be determined, it is important to understand how galaxy properties and correlations from the source survey (such as \GEMS) translate into the simulations \citep[see][]{2019A&A...624A..92K, SKiLLS}. We stress again that our simulations described above are only created for the purpose of comparing runtime improvements of different noise cancellation approaches. 
        
\begin{figure}
            \resizebox{0.49\hsize}{!}{\includegraphics{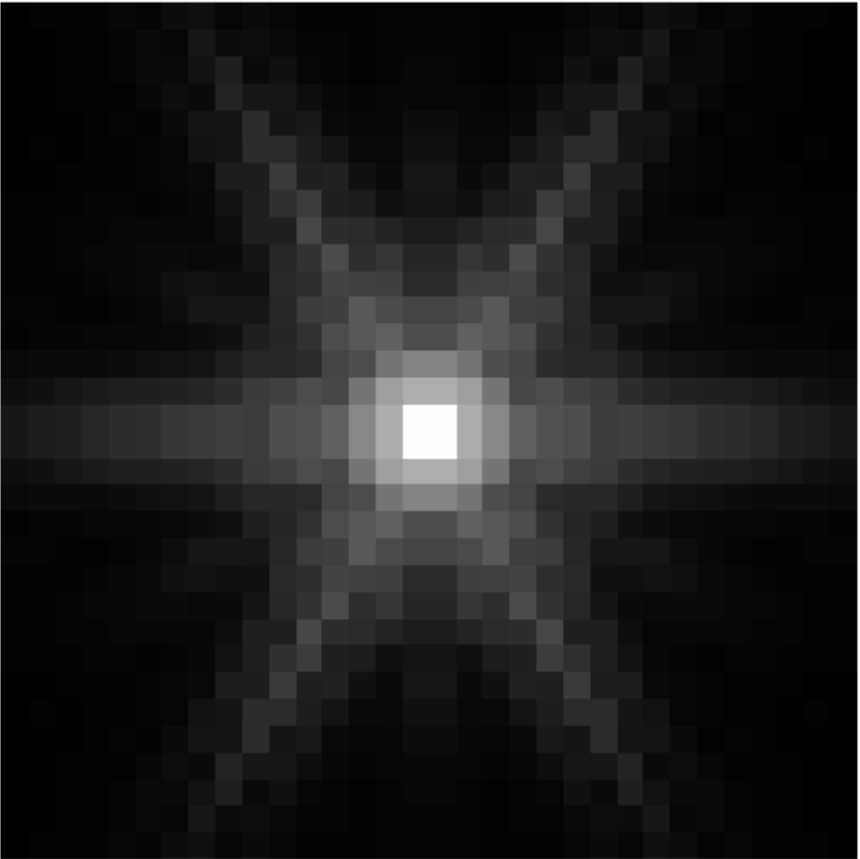}}
            \resizebox{0.49\hsize}{!}{\includegraphics{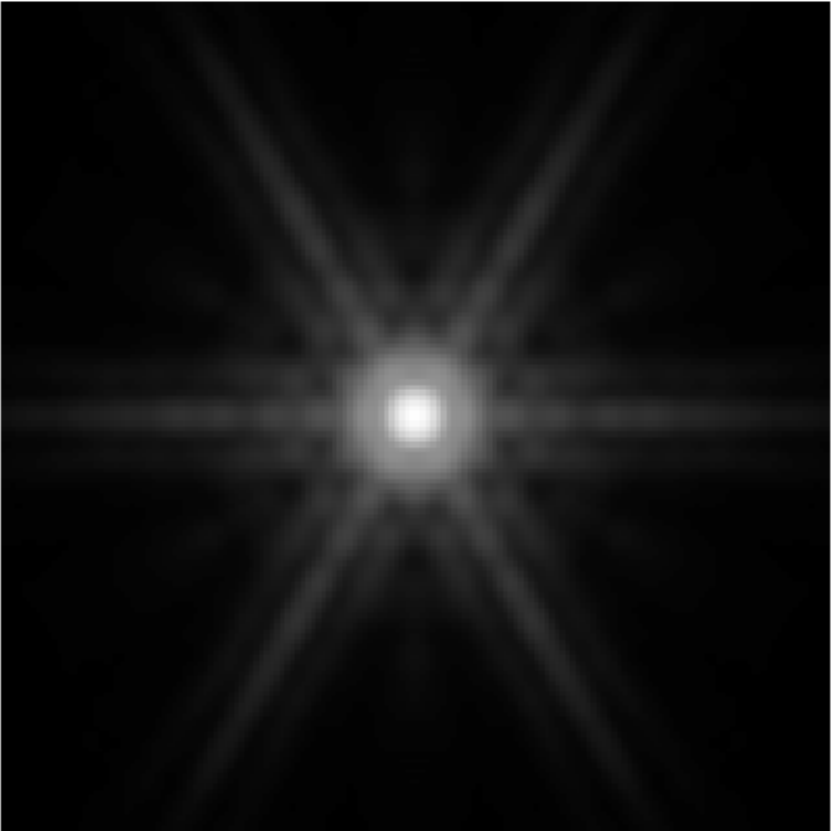}}
            \caption{Simulated PSF shown with a logarithmic grayscale. In the left panel, the PSF is drawn on the native $0 \farcs 1$ scale of \Euclid on a $32\times 32$ pixel grid. In the right panel, the PSF is first convolved with a $0\farcs 1$ filter function and then drawn on a finer $160\times 160$ pixel grid with a pixel scale of $0\farcs 02$ as used for the measurements with a subsampling factor of five.} 
            \label{fig:PSF}
        \end{figure}

        The PSF for the simulations was kept constant and modelled to be \Euclid-like. Adapting to the VIS bandpass from $550$ to $\SI{900}{\nano\meter}$ \citep[see][]{Cropper_2016}, we generated several monochromatic PSFs within this bandpass using the \galsim optical PSF function. We then stacked those monochromatic PSFs to obtain a single PSF, which is representative of the VIS bandpass. This stack was built from a weighted sum of the individual PSFs, where the weights were computed from a modelled Vega spectrum taken from \citet{Vega_Spectrum}. The telescope properties required for the optical PSF function were taken from \citet{Tewes_2019}. In Fig.~\ref{fig:PSF} the combined PSF can be seen on both the native pixel grid of \Euclid and also on the finer grid used for subsampled galaxy images.  We systematically simulated the galaxy images using the VIS native pixel scale. However to improve the reliability of the \galsim \KSB implementation on undersampled galaxies, we then subsampled all images with a factor five by subdividing each native pixel into $5\times 5$ subpixels before applying \KSB. Hereby the flux of one pixel is distributed evenly among $5\times 5$ new pixels and therefore the total number of image pixels is increased by a factor of $25$.\footnote{This manipulation of the pixel data is commonly used for \KSB \citep[see e.g.][]{Erben, 2006MNRAS.368.1323H, beatriz} and can also be applied to real data.}  
        
        For the determination of the additive bias $c$, we needed to make sure that the input ellipticities vanish on average. We introduced a shape noise cancellation in the input catalog to ensure that this holds.
        Since we drew an ellipticity and an orientation angle, we could introduce this by drawing a galaxy with angle $\alpha$ and the same galaxy with the same ellipticity but angle $\alpha + \pi /2$ once more. We did this for the response method on a grid and also on random positions. In the case of random positions, we used the same galaxies turned by $90$ degrees on new random positions in consecutive images. That way the shape is not only cancelled in the whole population of galaxies but also for each galaxy of the input catalog itself. Following this setup, we are able to estimate the additive bias without impacting the estimate of the multiplicative bias. 
        
        \subsubsection{Grid-based simulations}
            \begin{figure}
                \includegraphics[width=0.495\hsize]{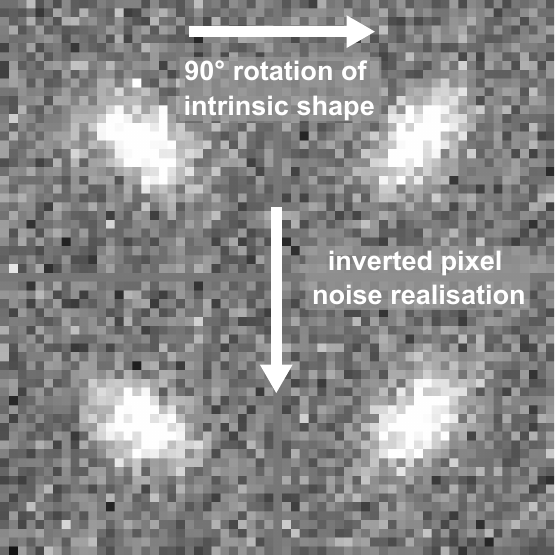}
                \includegraphics[width=0.495\hsize]{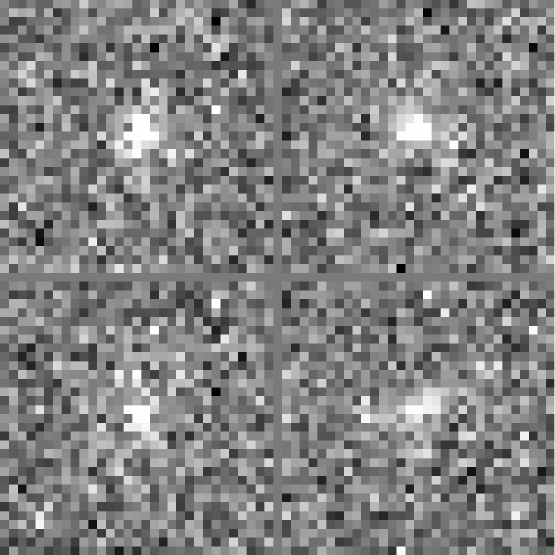}
                \caption{Examples for the grid-based simulations, where both cancellations are used. For better identification of individual pixels, galaxies are drawn on $32\times32$ pixel stamps here. In the horizontal direction shape noise cancellation is applied by rotating the galaxy by 90 degrees. Vertically pixel noise cancellation is implemented. As discussed in Sect.~\ref{sec:pixel_noise_cancellation}, this only approximately corresponds to switching the sign of the noise. The left panel shows a bright and extended galaxy for a better visual impression, while the right panel shows a more typical galaxy in the GEMS catalog.}
                \label{fig:grid_based_example}
            \end{figure}    
            In the grid-based simulations, we built stamps of $64\times 64$ pixels in size with isolated galaxies. Apart from a randomly distributed sub-pixel shift, these galaxies are centred within the stamp. No detection step is needed for the grid-based simulations. 
            However for more realism we applied a selection based on a measured signal-to-noise ratio larger than ten using the definition from \citet{Tewes_2019}.
            We subsampled each stamp by a factor of five and then ran \KSB via the \galsim shear estimate function directly on the subsampled stamp. Stamps for which the \KSB measurement fails were not considered for later analysis. We do not require completeness for the cancellations. If not all versions belonging to a cancellation could be measured, we still kept the ones that were measured successfully. Otherwise the selection bias would be artificially suppressed by the cancellations.
            
            \paragraph{Fit method}
                Depending on the cancellation method used, each galaxy has a certain number of stamps. For our shape noise cancellation, we made use of two stamps and for each stamp with added noise there was one with noise inverted. That way we ended up with four stamps per galaxy if both cancellations were applied. Two practical examples can be seen in Fig.~\ref{fig:grid_based_example}. Typically galaxies are small as in the right panel. In the left panel, the shape noise cancellation is easier to identify. Such panels consisting of several stamps were then built for 20 different shears evenly spaced from $-0.1$ to $0.1$ and for each shear a different galaxy sample was used. 
            
\begin{figure*}
                    \centering
                    \includegraphics[width=17cm]{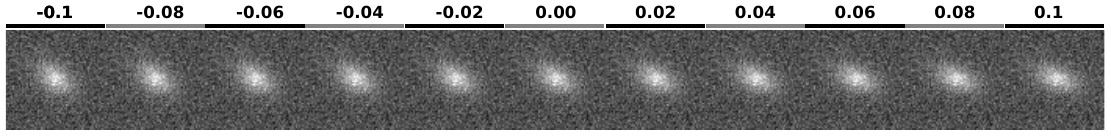}
                    \caption{One panel for the response method expanded over the full shear range of the fit methods from $-0.1$ to $0.1$. The legend above indicates the applied shear $g_1$ for each stamp. The shear difference $\Delta g$ is $0.02$ from stamp to stamp. One can observe the virtually identical noise pattern in each stamp, which only differs due to the Poisson noise of the galaxy itself as described in Sect.~\ref{sec:3_rm}.}
                    \label{fig:Pujol_example}
                \end{figure*}

            \paragraph{Response method}
                For the response method, one panel consists of the same galaxy sheared $n$ times. The original method uses $n=2$, but we extended this to $n=11$ if the larger shear interval was needed. Also a random sub-pixel shift was applied to the galaxies, but this sub-pixel shift is the same in one specific panel. In Fig.~\ref{fig:Pujol_example} an example for the $n=11$ case can be seen. The original method corresponds to only the $|g|=0.02$ stamps, where the shear difference is barely notable by eye. 

        \subsubsection{Galaxies at random positions}
            The second experiment consisted of $4000 \times 4000$ pixel-wide scenes containing galaxies from the \GEMS catalog with $m < 26.5$, where $m$ denotes the magnitude in the F606W filter of the HST. This magnitude cut was chosen such that the \GEMS catalog is still complete. As shown in \citet{2017MNRAS.468.3295H} and more recently in \citet{RefId0}, one has to include galaxies as faint as magnitude 29 to obtain accurate bias estimates because the estimate is affected by undetected blends. Since we are mainly interested in uncertainties of biases rather than their absolute value, we did not include the faintest galaxies to save computing time. 
            
\begin{figure}
                \resizebox{\hsize}{!}{\includegraphics{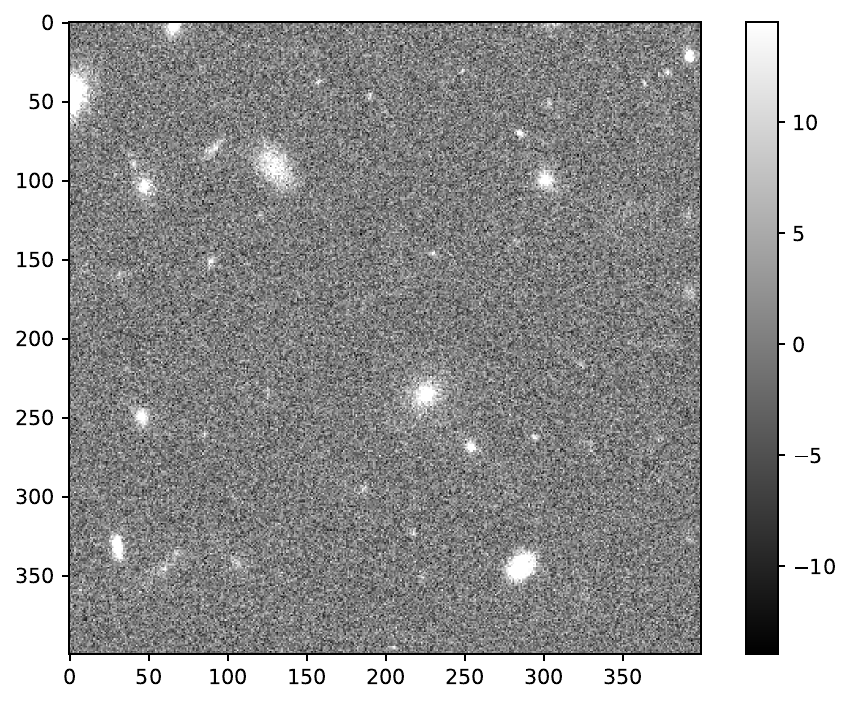}}
                \caption{Cutout of $400\times 400$ pixels from an exemplary larger scene. All galaxies visible here do have the same constant shear applied. The scene contains galaxies down to $\SI{26.5}{\mag}$ and has a density of $\SI{30}{galaxies\,(\mathnormal{m}<\SI{24.5}{\mag})\per\sqarcmin}$ as expected for \Euclid.} 
                \label{fig:large_scene_example}
            \end{figure}

            As a first step, we added Poisson noise from the sky background and read-out noise to the empty image. Then galaxies with a constant shear were drawn again on $64 \times 64$ pixel-wide stamps and only Poisson noise due to their own flux was added. These stamps were then added at random positions in the large scene. As a result, some regions might contain many blends, while others are less dense. If one stamp reaches above the margins of the large image, only the overlap is added. One example of a cut-out from one of the scenes generated in this way is shown in Fig.~\ref{fig:large_scene_example}. 
            We generated several new realisations of the random positions for each constant input shear. Hence for our final simulation run described in Table~\ref{tab:eff_rp}, we used 2800 $(=140\times 20)$ different random position realisations for the fit method. Statistical fluctuations of the amount of blending due to the specific realisation of the random positions are therefore suppressed. The mean density is chosen as $\SI{30}{galaxies\per\sqarcmin}$ brighter than magnitude $24.5$ to match the expectations of \Euclid. 
            
            Once generated these scenes were analysed with \SExtractor to generate detection catalogs. We used the default settings of \SExtractor (Version 2.25.0)\footnote{Essential settings are: $\texttt{DETECT\_MINAREA} = 5$, $\texttt{DETECT\_THRESH}=1.5$, $\texttt{DEBLEND\_NTHRESH}=32$, $\texttt{DEBLEND\_MINCONT}=0.005$}. The positions detected in this manner were then used to extract $64\times 64$ pixel large stamps. The extracted stamp's size was chosen so that it is large enough to cover the large galaxies and small enough to minimise the impact of blending. Also this stamp size makes comparing the results to the grid-based results easier. From this point on, the procedure is the same as for the grid-based simulations including selecting S/N ten and above. The detection steps and the blends might of course change the biases. Using this configuration of \SExtractor and the extraction flags one and two, which indicate an impact by neighbouring objects, we find a blending fraction of $2.5\%$ for the complete sample of galaxies. Here we defined the blending fraction as the ratio of detected objects, which had either one or both of the extraction flags raised, to the total number of detected objects. \citet{blending_synergies} find the blending fraction to be about $10\%$ for galaxies brighter than 26 mag in the \Euclid VIS band. Using their alternative definition of blending defined as an overlap of two or more Kron-ellipses, we find a blending fraction of $8.1\%$ in our random position simulations. Just as the authors we used apertures with a size of 2.5 times the Kron-radius, which captures about $90\%$ of the light from the galaxy with a slight dependence on the Sérsic index. As these two definitions of blending fractions can be easily impacted by the \SExtractor configuration and the simulation details, we validate our blending fractions against an alternative set of simulations that includes clustering, as described in Sect.~\ref{subsec:Flagship}. 
            
\begin{figure*}
                    \centering
                    \includegraphics[width=60mm]{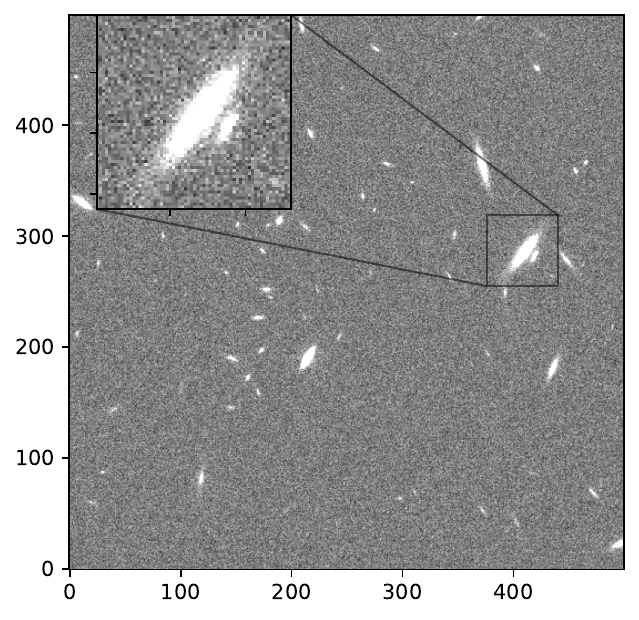}
                    \includegraphics[width=60mm]{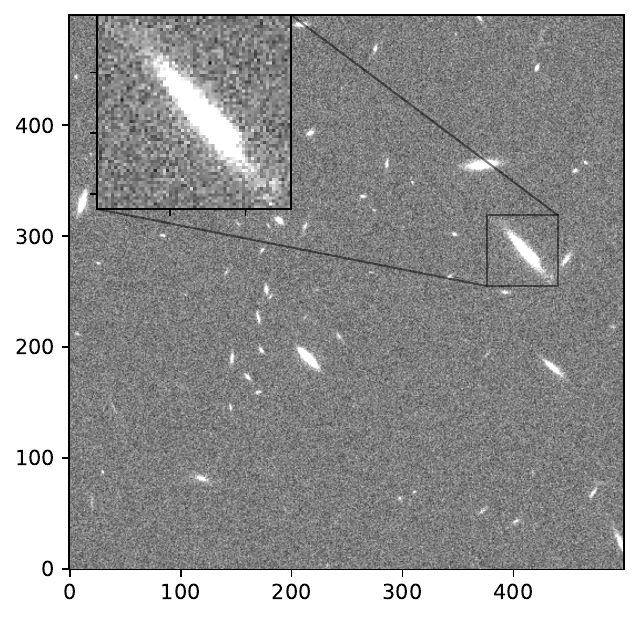}
                    \includegraphics[width=60mm]{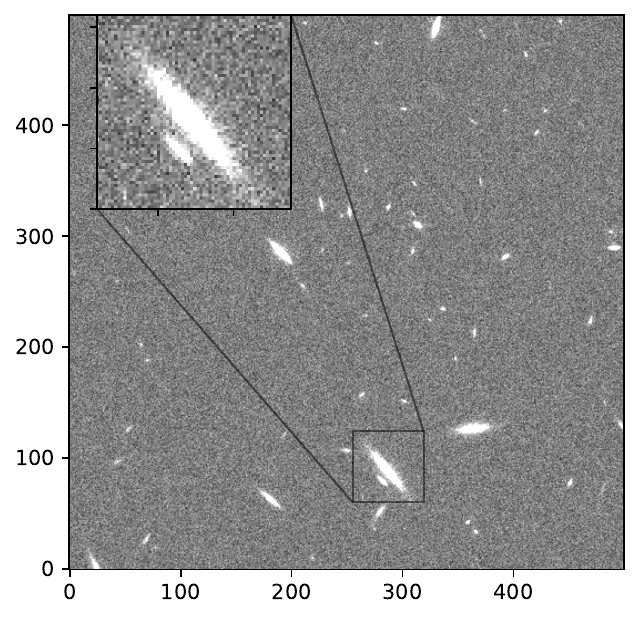}
                    \caption{Options for the shape noise cancellation on random positions. The left panel is shown for reference. The middle panel shows the shape noise cancellation with constant positions but rotated galaxies. The right panel implements the cancellation by rotating the whole scene before shearing by 90 degrees. The zoomed frames show the same pair of close-by galaxies in each version. The figure illustrates the benefit of the global cancellation, which keeps the galaxies distinct, while the local cancellation creates a blend.}
                    \label{fig:shape_noise_cancel_options}
                \end{figure*}

            \paragraph{Fit method}
                We generated up to four versions of the same large scene to use the different cancellation methods. We generated a second version of the scene with the same background noise for shape noise cancellation. The cancellation can then be implemented in different ways. One option is to keep the position of the galaxy centres constant and rotate them by 90 degrees. In the following, this is referred to as local shape noise cancellation. Still rotating the galaxies on fixed positions changes the relative blending. Therefore shape noise cancellation is likely to be less efficient. Another option is to rotate the whole scene before applying shear. We refer to this as global shape noise cancellation in the following. Thus the positions and the galaxies themselves are rotated by 90 degrees. That way the relative blending stays the same apart from the differences due to the shear. The blending level is given by the reference scene and is therefore comparable between the two cancellation approaches. Still the benefit of the global cancellation is that it preserves the relative alignment of the galaxies. Thus two blended galaxies in the reference frame stay blended in the rotated version, while isolated galaxies stay isolated. The local cancellation creates new blends in the rotated version while de-blending others. As a result, the shape noise cancellation is destroyed for more galaxies than it is the case for the global cancellation. The change in blending fraction compared to the reference scene is a noisy quantity for both types of shape noise cancellation, because we draw a new pixel noise realisation for the two versions of the scene. Since we use many different realisations of random positions, the effect of slightly varying blending fractions in the different versions of a scene is negligible. A drawback of the global cancellation is that it suffers from spatial variations of detector effects. Spatial variations of the quantum efficiency, the PSF, and other effects, like bad pixels would also have to be rotated. In particular a global cancellation with variable shears in the field likely causes issues and requires further exploration. We compare both of these approaches. An example of the difference between the two options can be seen in Fig.~\ref{fig:shape_noise_cancel_options}. To use both cancellations discussed in this paper, we generated one additional scene for each of the two scenes used for shape noise cancellation. These two additional scenes carry the same noise as their respective partner, except that it is inverted instead of added. Thus, one run of the fit method with both cancellations consists of $20\,(\text{different shears}) \times 4\,(\text{scenes per shear})$ $4000\times 4000$ pixel images. Our analysis, later on, is based on several of such runs combined. 
                
            \paragraph{Response method}
                For the response method, we need to generate two or eleven (depending on the shear interval of interest) differently sheared versions of the same scene. These scenes are identical in background noise and read-out noise and thus solely differ in the Poisson noise of the galaxies' light distribution. This was again generated from the same seed utilising the inverse transform sampling. One run here consists of a compilation of almost the same scene but slightly different shear. This can also be repeated several times for better statistics.

    \subsection{Definition of runtime}\label{subsec:runtime}
        The comparison of different methods to estimate the shear biases must be based on a definition of efficiency. A good proxy for efficiency is the time needed by the simulations to end up at a certain level of uncertainty. As the runtime of a simulation depends of course on the exact implementation (e.g. parallelisation) and the available number of CPU cores, a generic method to compare runtimes is useful. For our setup, the largest contributions to the overall runtime come from
        \begin{itemize}
            \item the drawing of each stamp (which includes the convolution with the PSF);
            \item the \KSB measurement;
            \item the noise generation (dominated by the inverse transform sampling for Poisson noise);
            \item and the generation of a $4000\times 4000$ pixel image from individual stamps.
        \end{itemize}
        The latter is only relevant for the simulations with galaxies at random positions. All the components were then timed many times and the resulting average was used to define relative runtimes in units of one convolution. On a particular single-CPU setup, we find that one convolution takes on average $\SI{0.14}{\second}$, one measurement takes $\SI{0.017}{\second}$, one noise generation takes $\SI{0.013}{\second}$, and building one large image takes $\SI{19.4}{\second}$. We normalised these times to one convolution and approximated from the times given above that noise generation and measurement of one stamp take $1/4$ the time of a convolution and building a large image takes $140$ times longer than a convolution. Expressing the latter as an image assembly time per galaxy yields $3\%$ of a convolution per galaxy, which shows that this step is very efficient. With the relative runtimes it is possible to calculate theoretical runtimes only depending on the number of convolutions, the number of \KSB measurements, and the number of $4000\times 4000$ pixel images a simulation run needs. In the following, all runtimes refer to this definition of a theoretical runtime. We want to emphasise that even this theoretical runtime only holds for our specific setup, as for example the stamp size changes the time \KSB takes compared to a convolution. Still the potential efficiency improvement for most of the methods is insensitive to the exact runtime differences between the main contributors listed above because most of the methods require all of the steps from the convolution to the ellipticity measurement. Adding shape noise cancellation for example always takes twice as long as using no cancellation since the same steps are also required for the rotated galaxy. Only the improvement of pixel noise cancellation is sensitive to the runtime difference between a convolution and the other contributors since this cancellation does not require an additional convolution. As long as the convolution with the PSF is the dominant part of the runtime, pixel noise cancellation has this advantage compared to the other methods.

\section{Uncertainty estimates}\label{sec:5}
    To assess the efficiency of different methods later on, the uncertainties on $\mu$ and $c$ need to be defined. This definition can become non-trivial depending on the method used and we discuss our treatment in the following. 
    \subsection{Grid-based simulations}
        
        The gridded setup makes it possible to identify a measurement with an input galaxy unequivocally. This simple mapping between measurement and input galaxy has some advantages in uncertainty determination. 
        \subsubsection{Fit method}
            We determined uncertainties for each point of the fit method using the bootstrap method generating $1000$ samples. The bootstrap was done over the galaxy population and not over the individual measurements to account for the purposely introduced correlation due to the cancellation. Afterwards $\mu$ and its respective $1\,\sigma$ uncertainty were determined using an implementation of the  Levenberg--Marquardt algorithm \citep{lm-algorithm}, which minimises the sum of squared residuals for (non-)linear functions. We judge the goodness of fit by utilising the $\chi^2$ statistic. Hereby $\chi^2$ is given as 
            \begin{equation}
                \chi^2 = \sum_{i=1}^{N} \biggl(\frac{x_{i}-\mu_{i}}{\sigma_{i}}\biggr)^2\,,
            \end{equation}
            where $N$ denotes the number of measurements, $\mu_{i}$ is the expectation value at position $i$ given by the model to fit, and $\sigma_{i}$ is the uncertainty of the measurement $i$. Further $\chi^2_\mathrm{red}$ is defined as 
            \begin{equation*}
                \chi^2_\mathrm{red} = \frac{\chi^2}{N-M}\,,
            \end{equation*}
            where $M$ denotes the number of fit parameters. The expectation value of this $\chi^2_\mathrm{red}$ for a good fit is one. We use the estimated uncertainties from the bootstrapping directly for the fitting and do therefore not enforce a rescaling of the $\sigma_i$ to yield a $\chi^2_\mathrm{red}$ of one. 
            
            This kind of fitting with absolute error bars is also consistent with a fitting, where the given error bars are only used for weighting. Thus estimating the uncertainties via bootstrapping for each data point is robust. Additionally we tested that an estimate with MCMC allowing for a constant fraction, by which we over- or underestimated our error bars, leads to consistent results. The uncertainty estimate is consistent in all these cases, so we are very confident in their value. In the following, we use the bootstrapping ansatz as this can be consistently done for all the methods.
        
        \subsubsection{Response method}\label{subsec:response_method}
            The response method is treated similarly. Again we can bootstrap over the galaxy population, where one galaxy might have two or eleven associated measurements depending on the chosen interval. The average response was calculated from Eq.~\eqref{eq:pujol_33}. Each galaxy has then an $e_{\mathrm{obs}}^+$ and an $e_{\mathrm{obs}}^-$ assigned to it. Then we can bootstrap those simultaneously by drawing galaxies (e.g. if a galaxy is included in the $\langle e_\mathrm{obs}^+ \rangle$ calculation then it is also included in the $\langle e_\mathrm{obs}^-\rangle$ calculation) and build responses from these samples. The standard deviation of those responses gives our uncertainty. 
            
            In the case of eleven versions of the same galaxy, there are two ways to estimate the multiplicative bias. One is to assign all but the version with the smallest shear to the $e_\mathrm{obs}^+$ and all but the version with the largest shear to the $e_\mathrm{obs}^-$. Then one can proceed using Eq.~\eqref{eq:pujol_33} again. The problem here is that the inner nine versions are included in both $e_\mathrm{obs}$ and therefore get a too large weight. This problem can be solved by determining the bias using a linear fit just like in the fit method. We simultaneously measured $\mu$ and $c$ by fitting the model from Eq.~\eqref{eq:bias_definition} to all galaxies and 11 input shears. The fitting was not done for each individual galaxy since that would be very noisy again. Since the same galaxies are included in each point used for the fit, we need to account for the correlation in the uncertainty estimate. This was done by bootstrapping the fit. We built bootstrap samples from the galaxies and fit for each bootstrap sample. The standard deviation of the fit parameters is taken as an uncertainty estimate. Effectively the response method is equivalent to a fit method with the same galaxies and noise for each input shear. We used this method to estimate the bias and its uncertainty for the response method on the large shear interval. 
            
    \subsection{Random position simulations}
        Defining uncertainties on the randomly positioned galaxy simulations has to be done differently than before since it is not possible anymore to identify complete cancellations or responses of individual galaxies. 
        
        \subsubsection{Fit method}
            For the fit method, we simulated several scenes per shear each giving a shear estimate by taking the mean of all observed ellipticities. Knowing also how many measurements went into the shear estimate of one scene, we could afterwards determine the combined shear estimate via a weighted average of all the individual estimates. The uncertainty of this combined estimate can then be obtained by bootstrapping over the individual estimates (again accounting for the weights). This uncertainty estimate is quite noisy for the fastest runtimes, where only a few scenes are being accounted for. Hence we left out the uncertainty estimates in our analysis, which were based on less than ten scenes. 
        
        \subsubsection{Response method}
            For the response method $\langle e^{\mathrm{obs, \,\pm}}_i\rangle$, where the index $i$ denotes the number of the run, can be found for each of the runs and later on combined into one larger $\langle e^{\mathrm{obs,\, \pm}}\rangle$ to build the response as described in Eq.~\eqref{eq:pujol_33}. Recall that one run refers to one compilation of scenes with slightly different shear, but essentially almost the same noise (so either a combination of two or eleven scenes). We used multiple runs for better statistics. The uncertainty was determined by bootstrapping $\langle e^{\mathrm{obs,+}}\rangle$ and $\langle e^{\mathrm{obs,-}}\rangle$ simultaneously to account for the correlation and building responses from these samples. To do so we drew random indices with repetition and then built the bootstrap samples by taking the $\langle e^{\mathrm{obs, \,\pm}}_i\rangle$ at the drawn indices to form new large $\langle e^{\mathrm{obs,\, \pm}}\rangle$ and build the response from these. The standard deviation of those responses is taken as the uncertainty. In the large shear interval with eleven versions we also employed the linear fit again and determined the uncertainty by bootstrapping the fit as described in Sect.~\ref{subsec:response_method}.

\section{Results}\label{sec:6}
    \subsection{Non-linearity of the shear measurement}
        \begin{figure}
            \resizebox{\hsize}{!}{\includegraphics{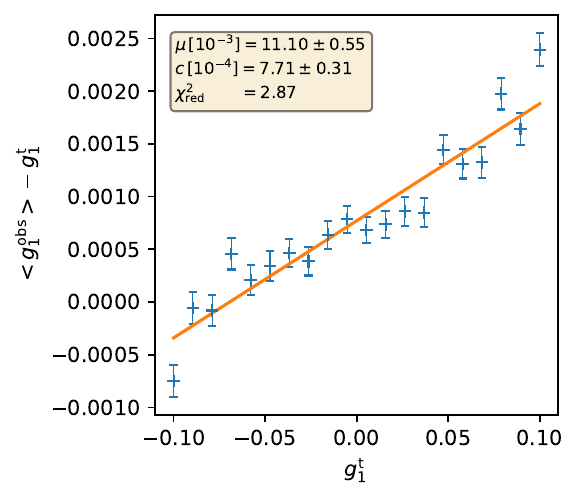}}
            \resizebox{\hsize}{!}{\includegraphics{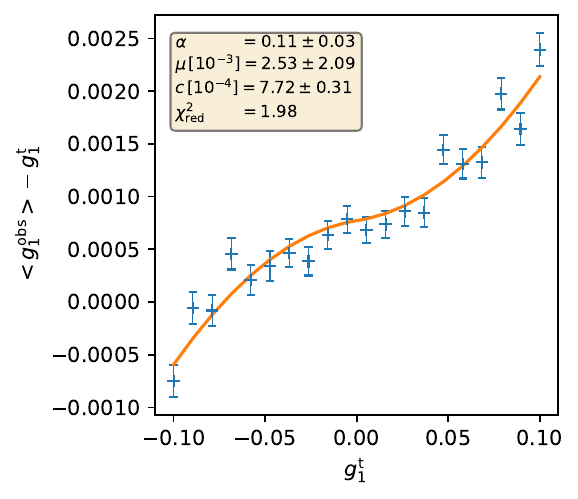}}
            \caption{Non-linearity in the shear measurement. The standard linear-fit method is shown in the upper panel, using gridded simulations with 160\,000 galaxies. The orange line indicates the fit. On the $y$-axis the difference between the average observed -- or measured -- shear and the true input shear is plotted. The lower panel shows the same data, but this time allowing for a quadratic term in the fit (fitting the function shown in Eq.~\ref{eq:quadratic_fit}). Here the additional parameter $\alpha$ describes the coefficient of the quadratic term. The linear fit has 18 degrees of freedom, while the quadratic fit has 17.}
            \label{fig:non-linearity}
        \end{figure}
        In both, grid-based and non-grid-based simulations, the original response method using only small shears yields different results than the fit method spanning a larger shear interval. The non-linearity of the shear measurement is clearly evident in Fig.~\ref{fig:non-linearity}. Fitting the same data allowing for an additional antisymmetrised quadratic term, describes the visible behaviour better and yields a better\footnote{We find that allowing for an additional $g_2$ component of the shear, the $\chi^2_\mathrm{red}$ can be further optimised towards unity. For each galaxy, we randomly assign an additional $g_2$ component binned similarly to the $g_1$ component. This increases the input ellipticity variance along the $g_2$ axis, yielding a better $\chi^2_\mathrm{red}$.} $\chi^2_\mathrm{red}$. To describe the antisymmetric behaviour we fit the function
        \begin{equation}\label{eq:quadratic_fit}
            \langle g_1^\mathrm{obs} \rangle - g_1^\mathrm{t} = \alpha\, |g_1^\mathrm{t}|\, g_1^\mathrm{t} + \mu\, g_1^\mathrm{t} +c\,.
        \end{equation}
        The non-linearity can also be seen in Fig.~\ref{fig:method_comparison_grid}, where we compare the multiplicative bias values of the different methods and specifically study the small shear interval as well by conducting the fit with 20 points in the interval $[-0.02, 0.02]$ (see the point \enquote{both 0.02}). This point contains three times more galaxies than the \enquote{both} point for the larger shear interval does. Still the uncertainty is way larger, illustrating the low efficiency of the fit method when using smaller shear ranges. Thus we need to conduct our accuracy comparison also dependent on the shear range. This behaviour is expected because the fit methods in the simplest form aim to estimate the slope
        \begin{equation}
            m = \frac{\Delta y}{\Delta x} \,.
        \end{equation}
        Using error propagation, the uncertainty on $m$ also depends on the $\Delta x$. Intuitively the signal-to-noise ratio on a shear estimate is lower at smaller shears.
        
        This non-linearity explains why the original response method (i.e. without fit) is always slightly less biased in the result tables than the other methods. The overall reason for the non-linearity, on the other hand, is not well understood, but it is plausible that the success rate of a \KSB measurement depends on the shear. We find a slight tendency that the measurement success rate of \KSB decreases with increasing amplitude of the shear. Additionally the fraction of complete cancellations decreases for larger shear, while the fraction of incomplete cancellations increases in return. The effect of higher-order terms in the shear bias was also studied recently by \citet{kitching}. We find that including a quadratic term in the shear bias improves the $\chi^2_{\mathrm{red}}$ a lot as seen in Fig.~\ref{fig:non-linearity}. As \citet{kitching} discusses, this behaviour might be caused by a projection of the total $g_1$-$g_2$ plane, but a detailed analysis of this effect is beyond the scope of this paper. The non-linearity can be neglected as long as the shear range is not too large and the comparison of methods is made in the same interval. For the comparison done in this paper it is a substantial effect, which we accounted for by expanding the shear range of the response method up to the range of the fit method.

    \subsection{Compatibility of methods}\label{sec:compatibility}
        Before we can compare the efficiency of different methods, we need to show that they can provide similar results in terms of $\mu$ and $c$. We do not want to have a method that yields smaller statistical errors but is not capable of providing the correct biases. We used the fit method without any cancellation as a benchmark for the comparison. The reference efficiency for all methods was therefore determined by looking only at fully independent versions of galaxies without any cancellations. In the case of Fig.~\ref{fig:grid_based_example}, this corresponds to considering only the top left version of all galaxies. For shape noise cancellation, we considered the top two versions and for both noise cancellations then all four. This was done analogously also for the simulation setup with larger scenes. Here we also took only one of the four different realisations of a scene into account for the reference efficiency. 
        \subsubsection{Grid simulations}
            \begin{table*}
                \caption{Method comparison on the grid.}
                \label{tab:mc_grid}
                \centering 
                \begin{tabular}{l c c c c c c} 
                \hline\hline 
                Method & $\mu$ & $\sigma_\mu$ & $c$ & $\sigma_c$ & simulated galaxies & relative\\
                    & $[10^{-3}]$ & $[10^{-3}]$ & $[10^{-4}]$ & $[10^{-4}]$ & $[10^6]$ & runtime\\
                \hline \Tstrut%
                No cancel & $12.85$ & 2.96 & 5.4 & 1.8 & 3.2 & 1.0\\
                Shape & $10.64$ & 0.86 & 7.8 & 0.5 & 6.4 & 2.0\\
                Both & $11.10$ & 0.55 & 7.7 & 0.3 & 12.8 & 2.4\\
                RM 0.1 (resp.) &  $8.74$ & 0.78 & 8.3 & 1.2 & 10.6 & 3.3\\
                RM 0.1 (fit) &  $11.34$ & 0.69 & 8.1 & 1.2 & 10.6 & 3.3\\
                Both 0.02 & $6.53$ & 1.40 & 7.0 & 0.2 & 38.4 & 7.5\\
                RM 0.02 & $6.84$ & 0.85 & 7.4 & 0.7 & 6.4 & 2.0\\
                
                \hline 
                \end{tabular}
                \tablefoot{We list the total number of simulated galaxies (including additional versions for the cancellations) for each method in the second last column. The relative runtime is always compared to the case without any cancellation. Only galaxies brighter than 24.5 magnitudes fulfilling the S/N larger than ten criterion are considered here. The RM in the first column denotes the response method. If the method is followed by a float number, this denotes the used shear interval. For the differentiation between response (resp.) and fit approach for RM 0.1 see Sect.~\ref{subsec:response_method}.}
            \end{table*}
            \begin{figure}
                \resizebox{\hsize}{!}{\includegraphics{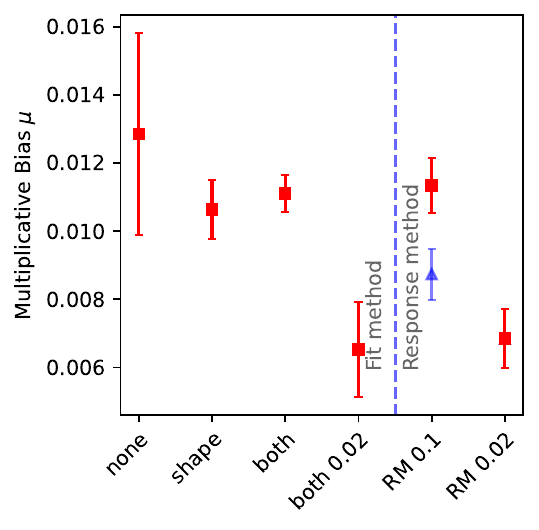}}
                \caption{Comparison of the multiplicative bias estimates from the different methods discussed in this paper on the grid. \enquote{Both 0.02} stands for the fit method using both cancellations, but in the smaller shear interval. The triangle symbol indicates the bias of the response method calculated without fitting. On the $x$-axis \enquote{RM} indicates estimates using the response method.}
                \label{fig:method_comparison_grid}
            \end{figure}
            The comparison is shown for the grid-based simulations in both Table~\ref{tab:mc_grid} and Fig.~\ref{fig:method_comparison_grid}. Note that we differentiate in both the table and the plot between the two ways to estimate the multiplicative bias for the response method in the large shear interval (see Sect.~\ref{subsec:response_method}). The direct approach using the responses is shown with a blue triangle, while the fitting approach is shown in red. The figure shows two important aspects of our study. Firstly using the same range of shears for the simulation leads to compatible results for every method. Hence we can compare the time it takes for a certain method to reach some accuracy. Secondly we can observe the non-linearity in the shear measurement that we discussed in the previous section. Changing the interval from $[-0.1, 0.1]$ to $[-0.02, 0.02]$ changes the multiplicative bias estimate. That is why we always denote the interval used for the response method in the following. The fit method always uses $[-0.1, 0.1]$ if not stated otherwise. 
            \begin{figure}
                \resizebox{\hsize}{!}{\includegraphics{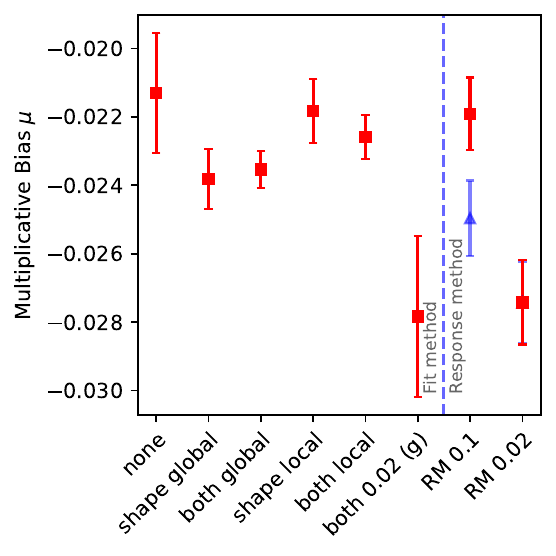}}
                \caption{Comparison of the multiplicative bias estimates from the different methods discussed in this paper for galaxies on random positions. The triangle symbol indicates the bias of the response method calculated without fitting. On the $x$-axis \enquote{RM} indicates estimates using the response method.}
                \label{fig:method_comparison_rp}
            \end{figure} 
        
        \subsubsection{Random positions}
            The same compatibility can be seen in the simulations with randomly positioned galaxies. This is shown in Table~\ref{tab:mc_no_grid} and Fig.~\ref{fig:method_comparison_rp}. The absolute value of the multiplicative bias becomes larger. Comparing the difference between grid and random positions with \citet[table 1]{RefId0}, we can attribute about a third of this shift to the inclusion of fainter galaxies and the other two-thirds to the additional detection step and in particular blending between brighter galaxies. The detection step, which we left out for the grid, leads to a detection bias for the fainter galaxies as shown in \citet{2021A&A...646A.124H}. Despite this shift, the methods are still compatible within their uncertainties. Only the methods using the smaller input shear interval deviate, which is likely due to the non-linearity again. For both types of simulations, we observe that the fit method and the response method yield consistent biases when the same shear intervals are employed and the bias of the response method is estimated by fitting (in the case of larger shears, see Sect.~\ref{subsec:response_method}). As for the grid simulations, biases are shifted towards more positive values (hence they are less negative) due to the non-linearity at larger shears. 
            \begin{table*}
                \caption{Method comparison on random positions.} 
                \label{tab:mc_no_grid} 
                \centering 
                \begin{tabular}{l c c c c c c} 
                \hline\hline 
                Method & $\mu$ & $\sigma_\mu$ & $c$ & $\sigma_c$ & simulated area & relative \\ 
                    & $[10^{-3}]$ & $[10^{-3}]$ & $[10^{-4}]$ & $[10^{-4}]$ & $[\si{deg\squared}]$ & runtime\\
                \hline \Tstrut%
                No cancel & $-21.30$ & 1.74 & 5.7 & 1.1 & 69.14 & 1.00\\ 
                Shape local & $-21.83$ & 0.93 & 7.2 & 0.6 & $2\times 34.57$ & 1.00\\
                Both local & $-22.59$ & 0.65 & 7.0 & 0.4 & $4\times 34.57$ & 1.22 \\
                Shape global & $-23.81$ & 0.88 & 7.8 & 0.5 & $2\times 34.57$ & 1.00 \\
                Both global & $-23.54$ & 0.54 & 8.0 & 0.3 & $4\times 34.57$ & 1.22 \\
                Both 0.02 (g) & $-27.83$ & 2.36 & 6.0 & 0.3 & $4\times 34.57$ & 1.22 \\ 
                RM 0.1 (resp.) & $-25.02$ & 1.10 & 10.5 & 2.1 & $11\times 4.94$ & 0.79 \\
                RM 0.1 (fit) & $-21.91$ & 1.06 & 10.1 & 2.1 & $11\times 4.94$ & 0.79 \\
                RM 0.02 & $-27.42$ & 1.19 & 5.5 & 0.9 & $2\times 24.69$ & 0.71 \\
                \hline 
                \end{tabular}
                \tablefoot{The relative runtime is given for this specific example where we used 11200 ($=140\times 20\times 4$) scenes for the fit method, 4400 ($=400\times 11$) scenes for the response method in the large shear interval, and 4000 ($=2000\times 2$) scenes for the response method in the small shear interval. It is always compared to the runtime of no cancellation. In the second last column, we list the simulated area that went into each method. Indicated is always the unique area multiplied by the required additional versions for each method. The RM in the first column denotes the response method. The estimate for Both 0.02 also used global (g) cancellation. If the method is followed by a float number, this denotes the used shear interval. For the differentiation between response (resp.) and fit approach for RM 0.1 see Sect.~\ref{subsec:response_method}.}
            \end{table*}

    \subsection{Uncertainty behaviour}\label{sec:uncertainty_behavior}
        In this section we want to study how the uncertainties develop using the different methods as a function of their runtimes. To do so, we assumed the simple dependence
        \begin{equation}
            \sigma_{\mu} = a\, t_\mathrm{run}^{-0.5} \,, 
        \end{equation}
        where $\sigma_{\mu}$ denotes the uncertainty on the multiplicative bias and $t_\mathrm{run}$ is the theoretical runtime defined previously. The same behaviour does also hold for the uncertainty of the additive bias. We expect this behaviour as the uncertainty scales with the inverse number of measurements and the number of measurements scales linearly with our runtime definition for a sufficiently large number of galaxies. 
        
        \subsubsection{Grid simulations}
\begin{figure*}
                \includegraphics{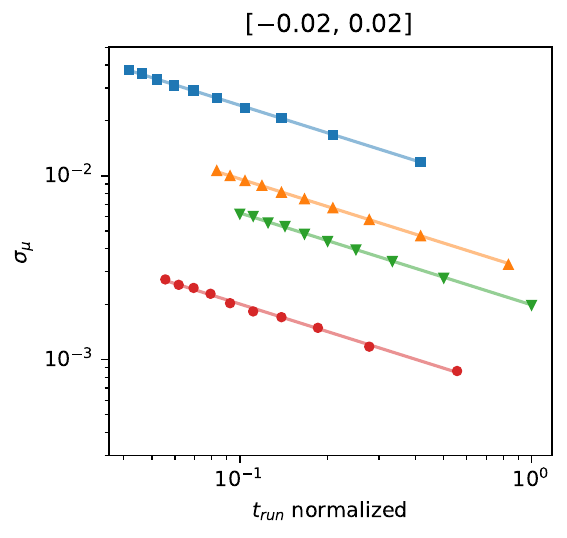}
                \includegraphics{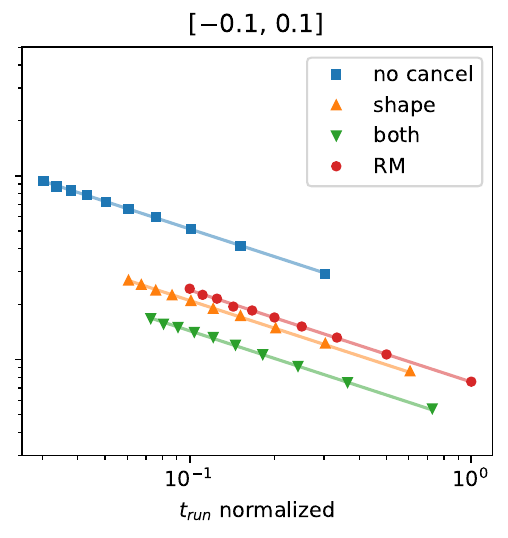}
                \caption{Uncertainty behaviour for the different methods for grid-based simulations. The runtime, which is used here, is always the theoretical runtime from Sect.~\ref{subsec:runtime} normalised to its maximum value in the respective figure. Here the large shear interval is normalised to a theoretical runtime of $13.2\times 10^6$ and the small shear interval to $14.4\times 10^6$, respectively. Solid lines show fits to the data described in Sect.~\ref{sec:uncertainty_behavior}. The left and right panels differ by the used shear interval, which is indicated in each column title.}
                \label{fig:uncertainty_behavior_grid}
            \end{figure*}

            On a grid, the uncertainty behaviour can be seen in Fig.~\ref{fig:uncertainty_behavior_grid}. 
            The runtime improvements deduced from the fitting results for the grid-based simulations can be seen in Table~\ref{tab:eff_grid}. The improvement in runtime can then be calculated by squaring the improvement of the fitted parameter $a$. Thus the definition of runtime improvement (RI) in our case is 
            \begin{equation}
                \mathrm{RI} = \paren{\frac{a_\mathrm{no\,cancel}}{a_i}}^2\,, 
            \end{equation}
            where $i$ denotes the method used for comparison. The table adds the total galaxy number improvement (GI). This quantity describes how many fewer galaxies need to be simulated with the respective method to reach the same precision as no cancellation. The total number includes all possible versions for the different cancellation methods. Thus using for example shape noise cancellation and pixel noise cancellation requires only a quarter of the total number of galaxies to be unique. The GI is independent of the runtime difference between convolution and other contributors to the runtime that are described in Sect.~\ref{subsec:runtime}. Therefore the GI can be used as a lower limit of the efficiency improvement if different shape measurement algorithms than \KSB are considered. For every method but both cancellations, RI and GI are the same. 
            
            For the multiplicative bias, we find that adding pixel noise cancellation on top of shape noise cancellation can reduce the runtime by another factor of about 2 compared to only shape noise cancellation. The improvement of the response method compared to no cancellation depends on the shear interval used. This is because the fit method uncertainty depends on the shear interval used, while the response method uncertainty is largely insensitive to it. In the small shear interval, it can provide an additional factor of $10$ improvement in runtime compared to both cancellations in the fit. \citetalias{Pujol_2018} find that the number of images can be reduced with the response method by a factor of 82 compared to shape noise cancellation and by a factor of 650 compared to no cancellation (assuming two sheared versions for the response method). They made the comparison in terms of simulated images and not in terms of runtime. Assuming that both are related linearly, we find a factor of $23$ in a number of images compared to shape noise cancellation and a factor of $145$ compared to no cancellation. Thus we do not exactly reproduce the values they found. Considering that we are using a completely different estimation of the uncertainties and a more realistic noise description, it is unsurprising that the estimated improvement deviates from their results. They also used a Gaussian distribution of the input shears with $\sigma=0.03$ such that the shear range is not entirely identical to our analysis. We also note that matching the shear interval of the response method to the fit method leads to a worse uncertainty behaviour again (compare RM in the different shear intervals). This is caused by the need to estimate multiple responses for the same galaxy. Hence there is more shot noise due to the galaxy population in the RM 0.1 method compared to RM 0.02 at the same runtime. 
            
            For the additive bias, the improvements of the fit method are always the same as they were for the multiplicative bias. Only the response method shows a significant difference. The response method is not effective at all currently when it comes to additive bias estimation. This originates from the need to simulate the same galaxy several times with slightly different shear. Since the additive bias is taken to be the mean of all observed ellipticities, the information gained from simulating the same galaxy multiple times is less than from simulating different galaxies. Thus the efficiency worsens when using eleven versions for the large shear interval. \citetalias{Pujol_2018} find larger improvement factors but took the additive bias as the mean of individual additive biases, which does not account for selection as discussed before. Thus the response method can not efficiently be used for additive bias estimation.  
            
            Nonetheless, we find two methods here that can reduce the multiplicative bias's runtime significantly compared to the commonly used shape noise cancellation. Using pixel noise cancellation on top of shape noise cancellation helps in every shear interval to reduce the runtime by at least a factor of 13, compared to a factor of 6 for shape noise cancellation only. In a large shear interval, the response method can not yield more improvement than both cancellations. In fact, its uncertainty behaviour with runtime is worse than that for both cancellations. But in a smaller shear interval, the response method improves the runtime by more than one order of magnitude. This can be very useful for redshift-dependent blending as discussed in \citet{SKiLLS} since the additive bias can also be calibrated empirically \citep[see][]{2021A&A...656A.135H}. Since galaxies are not on a grid in the sky, we also check what this behaviour looks like for randomly positioned galaxies. 
            
            \begin{table}
                \caption{Efficiency comparison for the grid-based simulations.}
                \label{tab:eff_grid}
                \centering 
                \begin{tabular}{l c c c c c c} 
                \hline\hline 
                Method & \multicolumn{3}{c}{$\mu$-bias} & \multicolumn{3}{c}{$c$-bias} \Tstrut\\
                & RI & $\sigma_\mathrm{RI}$ & GI & RI & $\sigma_\mathrm{RI}$ & GI \\
                \hline %
                \multicolumn{7}{c}{Shear interval $[-0.1, 0.1]$}\Tstrut\Bstrut \\
                \hline 
                Shape & 6.0 & 0.1 & 6.0 & 6.2 & 0.1 & 6.2\Tstrut\\
                Both & 12.8 & 0.2 & 7.7 & 13.9 & 0.3 & 8.3\\
                RM & 4.6 & 0.1 & 4.6 & 0.7 & 0.01 & 0.7 \\
                \hline 
                \multicolumn{7}{c}{Shear interval $[-0.02, 0.02]$}\Tstrut\Bstrut \\
                \hline 
                Shape & 6.3 & 0.1 & 6.3 & 6.3 & 0.1 & 6.3 \Tstrut\\
                Both & 14.9 & 0.3 & 8.9 & 14.9 & 0.3 & 9.0\\
                RM & 145.6 & 2.7 & 145.6 & 3.4 & 0.1 & 3.4\\
                \hline
                \end{tabular}
                \tablefoot{This table includes galaxies with input magnitudes brighter than 24.5 and a selection of S/N ten and above. RM stands for the response method, RI for runtime improvement, and GI for the total galaxy number improvement. The uncertainty for the runtime improvement is listed as $\sigma_\mathrm{RI}$.}
            \end{table}
        
        \subsubsection{Random positions}
\begin{figure*}
                \includegraphics{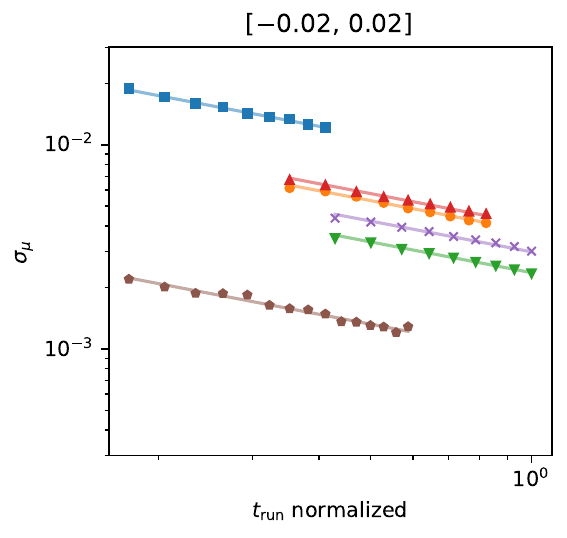}
                \includegraphics{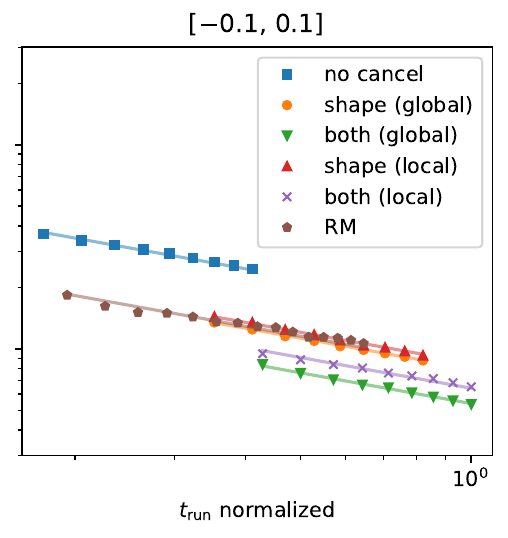}
                \caption{Uncertainty behaviour for the different methods for galaxies at random positions. Here both panels are normalised to a theoretical runtime of $46\times 10^6$. The figure is otherwise similar to Fig.~\ref{fig:uncertainty_behavior_grid}.}
                \label{fig:uncertainty_behavior_rp}
            \end{figure*}

            The uncertainty behaviour for galaxies at random positions is shown in Fig.~\ref{fig:uncertainty_behavior_rp}. The general trend is the same as for the grid-based simulations. In larger shear intervals, the response method does not further improve the efficiency compared to using both cancellations. For small shear intervals, there is still a very significant improvement in runtime using the response method. This is also supported by the detailed fit results shown in Table~\ref{tab:eff_rp}. In this table, we added the area improvement (AI), which works analogously to the total galaxy number improvement (GI) for the grid-based simulations. It describes how much less area needs to be simulated with the respective method to reach the same precision as no cancellation. This area includes all possible versions for the different methods. Thus using for instance shape noise cancellation and pixel noise cancellation requires only a quarter of the total simulated area to be unique. Since it does not depend on the runtime differences between convolution and other contributors to the runtime that are described in Sect.~\ref{subsec:runtime}, it gives a lower limit for the efficiency improvement if other shape measurement methods than \KSB are considered. For all methods but both noise cancellations, RI and AI are the same. 
            
            In general, all methods are not as efficient anymore as they were on a grid. We attribute this to blending and the inclusion of fainter galaxies, but also due to the additional detection step using \SExtractor, which has been left out on a grid. Nonetheless the advantages of either adding pixel noise cancellation or even using the response method are still present for the multiplicative bias estimation. The efficiency of the additive bias estimation is also here very poor for the response method. Only the fit method can provide improvements for the additive bias estimation. We also see that the global shape noise cancellation (and especially the global shape noise cancellation with pixel noise cancellation) is more efficient than the local one. This is expected since the relative blending in the global case stays constant, as previously mentioned. Thus if the purpose of the simulation allows it, global cancellation should be used. If that is not possible, the local cancellation can provide a seven times faster bias estimation than without any cancellation. 
             
            \begin{table}
                \caption{Efficiency comparison for galaxies at random positions.} 
                \label{tab:eff_rp} 
                \centering 
                \begin{tabular}{l c c c c c c} 
                \hline\hline 
                Method & \multicolumn{3}{c}{$\mu$-bias} & \multicolumn{3}{c}{$c$-bias} \Tstrut\\
                & RI & $\sigma_\mathrm{RI}$ & AI & RI & $\sigma_\mathrm{RI}$ & AI\\
                \hline %
                \multicolumn{7}{c}{Shear interval $[-0.1, 0.1]$}\Tstrut\Bstrut \\ 
                \hline
                Shape local& 3.6 & 0.3 & 3.6 & 3.8 & 0.2 & 3.8\Tstrut\\
                Both local & 6.3 & 0.6 & 3.9 & 6.8 & 0.5 & 4.2\\
                Shape global & 3.8 & 0.3 & 3.8 & 4.1 & 0.3 & 4.1\\
                Both global & 8.4 & 0.6 & 5.1 & 9.5 & 0.6 & 5.8\\
                RM & 3.7 & 0.3 & 3.7 & 0.3 & 0.02 & 0.3\\
                \hline 
                \multicolumn{7}{c}{Shear interval $[-0.02, 0.02]$}\Tstrut\Bstrut \\
                \hline 
                Shape local & 3.7 & 0.2 & 3.7 & 3.8 & 0.2 & 3.8\Tstrut\\
                Both local & 6.9 & 0.3 & 4.2 & 7.1 & 0.4 & 4.3\\
                Shape global & 4.3 & 0.3 & 4.3 & 4.2 & 0.3 & 4.2 \\
                Both global & 10.9 & 1.1 & 6.6 & 10.5 & 0.7 & 6.4\\
                RM & 69.5 & 3.3 & 69.5 & 1.7 & 0.1 & 1.7\\
                \hline
                \end{tabular}
                \tablefoot{This table includes galaxies with input magnitudes brighter than 26.5, and the same S/N larger than ten selection. RM stands for the response method, RI for runtime improvement, and AI for area improvement. The uncertainty for the runtime improvement is listed as $\sigma_\mathrm{RI}$.}
            \end{table}
    
     \subsection{Binned comparison}
     
        In addition to looking at the uncertainty behaviour for realistic scenes, we can also study the behaviour as a function of magnitude. In this section, we repeat the analysis in bins of input magnitude. The runtime improvement is always defined compared to no cancellation in the same magnitude bin. We focus here on the results of the multiplicative bias. The analysis of the additive bias shows a similar qualitative behaviour and is shown in Appendix~\ref{sec:app_binned_improvements_c}. Additionally we can also compare the absolute biases in different magnitudes bins just like we did it for the whole sample in Sect.~\ref{sec:compatibility}. This is a further validity check for the methods that we present in Appendix~\ref{sec:app_absolute_biases}.
        
        \subsubsection{Grid simulations}
            \begin{figure*}
                \includegraphics{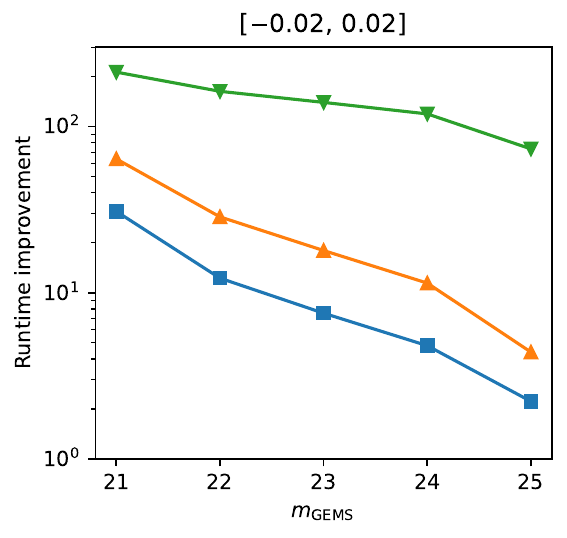}
                \includegraphics{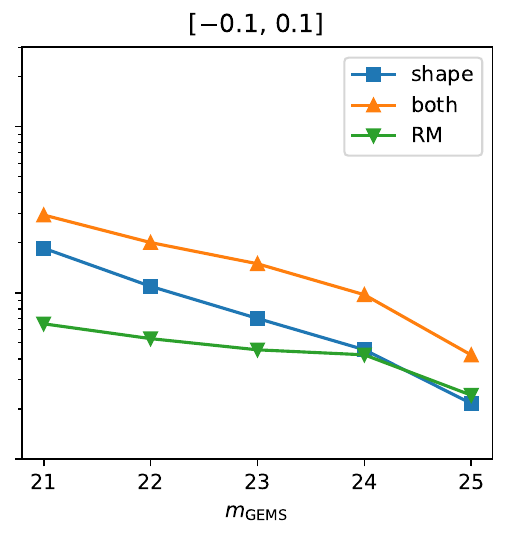}
                \caption{Magnitude-binned runtime improvement of the multiplicative bias for the grid-based simulations. The position of the points marks the center of each bin. The runtime improvement is always compared to the fit method without any cancellation. Error bars are smaller than the symbols and therefore omitted for better visibility.}
                \label{fig:binned_improvement_grid}
            \end{figure*} 
            The grid simulations make it trivial to bin in input magnitude. We extend the magnitude range from $20.5$ to $25.5$, which includes fainter galaxies than our previous grid analysis. This interval is binned in five bins with a width of one magnitude. The results can be seen in Fig.~\ref{fig:binned_improvement_grid}. We see the general trend of all methods becoming less effective for fainter galaxies. This is expected since signal-to-noise ratios are smaller for fainter galaxies and magnitudes correlate with the size of the galaxies.
            \paragraph{Fit method}
                The slope of this decrease with magnitude in efficiency is almost the same for the two cancellation methods so that using both cancellations always stays superior to using only shape noise cancellations. 
            \paragraph{Response method}
                Especially in the larger shear interval, we observe that the response method has a slower decline of the runtime improvement with magnitude than the fit method. While being the least efficient method for very bright galaxies, it becomes more efficient than shape noise cancellation for the faintest galaxies. This is probably due to the fact that pixel noise dominates the images of the faintest galaxies and the response method handles this kind of noise differently from the fit method. Hints of this flatter decrease can also be seen for the smaller shear interval, but the response method stays the most efficient method here anyway for all magnitudes. 
                
        \subsubsection{Random positions}
            \begin{figure*}
                \includegraphics{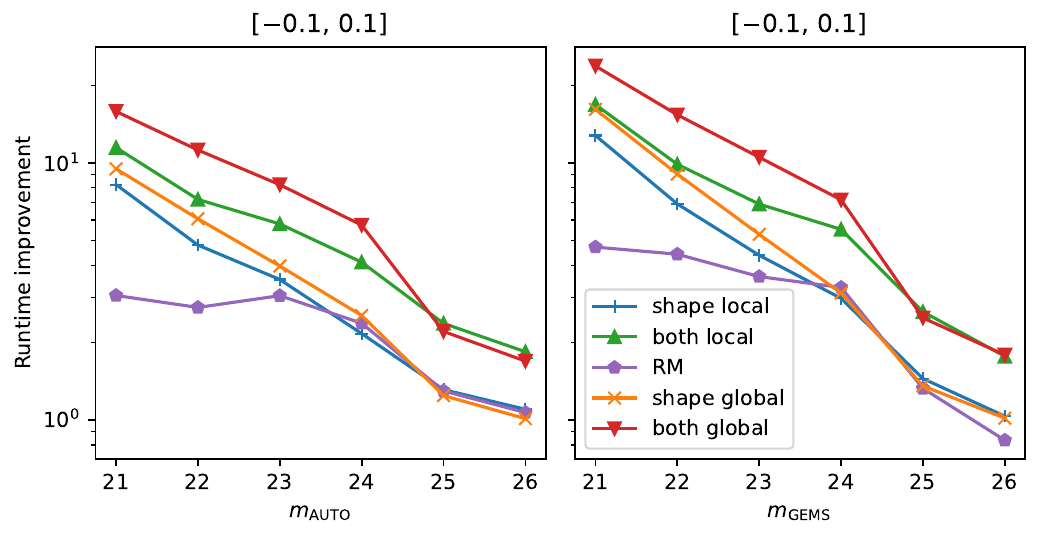}
                \caption{Magnitude-binned runtime improvement of the multiplicative bias for the galaxies on random positions in the large shear interval. In the left panel, the data is binned in $m_\mathrm{AUTO}$ from \SExtractor, while the input magnitude is used for the binning in the right panel to point out the differences. The position of the points marks the center of each bin. The runtime improvement is always compared to the fit method without any cancellation. For this figure no signal-to-noise cut is applied. Error bars are smaller than the symbols and therefore omitted for better visibility.}
                \label{fig:binned_improvement_rp_01}
            \end{figure*} 
            \begin{figure*}
                \includegraphics{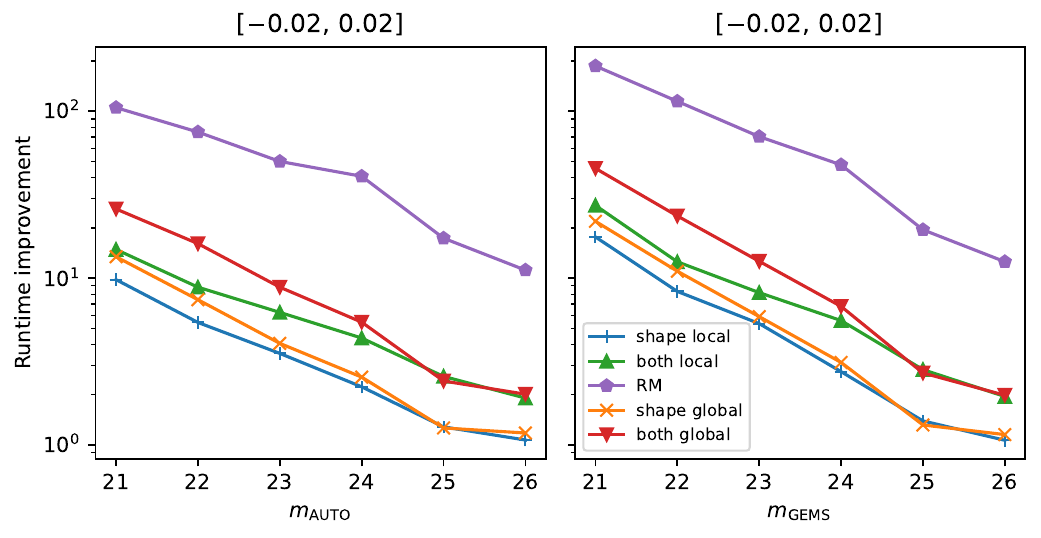}
                \caption{Magnitude-binned runtime improvement of the multiplicative bias for the galaxies at random positions in the small shear interval. In the left panel, the data is binned in $m_\mathrm{AUTO}$ from \SExtractor, while the input magnitude is used for the binning in the right panel to point out the differences. The position of the points marks the center of each bin. The runtime improvement is always compared to the fit method without any cancellation. For this figure, no signal-to-noise cut is applied. Error bars are smaller than the symbols and therefore omitted for better visibility.}
                \label{fig:binned_improvement_rp_002}
            \end{figure*} 
            On random positions, this behaviour looks different. Firstly matching between the detection catalog from \SExtractor and the input catalog is required to bin in input magnitudes. This position-based matching is of course not perfect, but at the implemented level of blending it works sufficiently well. It enables us to bin based on input magnitudes, which minimises magnitude-related selection biases. Still for real survey data only measured magnitudes are available. Therefore we also perform the same analysis with a binning based on the \SExtractor \texttt{MAG\_AUTO} magnitude. The magnitude range for these simulations spans from $20.5$ to $26.5$. We use six bins with each bin being one magnitude wide. The last bin from $\SI{25.5}{\mag}$ to $\SI{26.5}{\mag}$ is usually very noisy as only very few of these galaxies pass the signal-to-noise ratio cut. To show the behaviour at faint magnitudes more clearly, we removed the signal-to-noise ratio cut for the binned analysis of the random position simulations shown in Fig.~\ref{fig:binned_improvement_rp_01} (large shear interval) and Fig.~\ref{fig:binned_improvement_rp_002} (small shear interval). In general, the type of magnitude used to define the binning seems to have the largest impact on the brightest bin. Binning in \SExtractor's $m_\mathrm{AUTO}$ worsens the improvement in the brightest bin compared to the binning against input magnitudes. For fainter galaxies, the binning has less impact and the improvement factors agree again. Since the brightest bin contains the fewest galaxies, it is the most sensitive to wrong bin assignments. Especially when these wrong bin assignments destroy the cancellation by assigning only a part of the galaxies belonging to the cancellation to the wrong bin. Additionally blending scenarios can only be correctly binned in $m_\mathrm{AUTO}$. Our nearest neighbour assignment can only pick up the magnitude of one of the blending partners, which results in a fainter magnitude than the total magnitude of the blended object. Thus blended objects tend to be in the brighter bins in $m_\mathrm{AUTO}$. This wrong assignment of blended objects worsens the runtime improvement, because blending typically results in a noisier measurement. Furthermore, we observe that all methods again become less efficient for fainter galaxies, as seen on the grid. 
            
            In Fig.~\ref{fig:Fig15} the blending fractions in different magnitude bins can be seen. It becomes evident that the brightest bins are more affected by blending and therefore also by objects scattering up into brighter bins as described before. Still this only affects the exact binning. The total number of detected galaxies is nearly the same for both magnitude definitions. Thus the difference blending makes for the runtime improvement can only be quantified by comparing Tables~\ref{tab:eff_grid} and \ref{tab:eff_rp}. There one can read off that the amount of blending implemented in our simulations lowers the runtime improvements by at most a factor 1.5.  
            \paragraph{Fit method}
                For the fit methods, we observe that the global shape noise cancellations are always more effective than their local counterparts for magnitudes brighter than $24.5$. At fainter magnitudes, global and local cancellations have roughly the same efficiency. This behaviour is observed for both $m_\mathrm{AUTO}$ and $m_\mathrm{GEMS}$ binning, which is why it is likely not related to wrong bin assignments due to the blending. Shape noise cancellation is not significantly faster than no cancellation in the two faintest bins, making it irrelevant if this cancellation is performed globally or locally. Since both cancellations include shape noise cancellations, we observe the same behaviour. Still using both cancellations is in every magnitude bin more efficient than only shape noise cancellation. This behaviour can be observed for both shear intervals.
              
            \paragraph{Response method}
                The response method on the large shear interval is the least effective method for the brightest galaxies. For magnitudes fainter than 23.5, it becomes about as effective as shape noise cancellation and also yields a similar performance at fainter magnitudes. This trend is also reflected in the combined results, where we see that the response method is less effective than both noise cancellations, but just as effective as only shape noise cancellations on this shear interval (see Table~\ref{tab:eff_rp}). In the smaller shear interval, the response method constantly dominates the runtime improvement over the fit method. 
                
    \subsection{Comparison with Flagship}\label{subsec:Flagship}
        Our simulation setup, which cuts the magnitude distribution at 26.5 mag and uses randomly positioned galaxies, is only a simplified picture of the expected images from \Euclid. To quantify the impact of more realistic clustering and the inclusion of fainter galaxies into the simulation, we made use of the Flagship simulation mock galaxy catalogue (Euclid Collaboration: Castander et al., in prep.), which was obtained from CosmoHub \citep{CosmoHub2, CosmoHub1}. The catalogue was generated by painting galaxies to an N-body simulation \citep{Flagship}, which only includes dark matter. This painting was done using a combination of a halo occupation model and abundance matching. The morphology of the galaxies was then modelled with a similar approach as followed in \citet{MICE}. We use from this catalogue both the galaxy positions on the sky and the galaxy morphology. Instead of single-Sérsic profiles, we adopted the double-Sérsic profile description available in the Flagship catalogue. Apart from the more realistic positions and the adapted morphology, we setup the simulations in the same way as the random position simulations. In particular the noise properties, the PSF, and the detection configuration of \SExtractor are the same.
        \subsubsection{Blending fractions}
\begin{figure}
                \centering
                \includegraphics{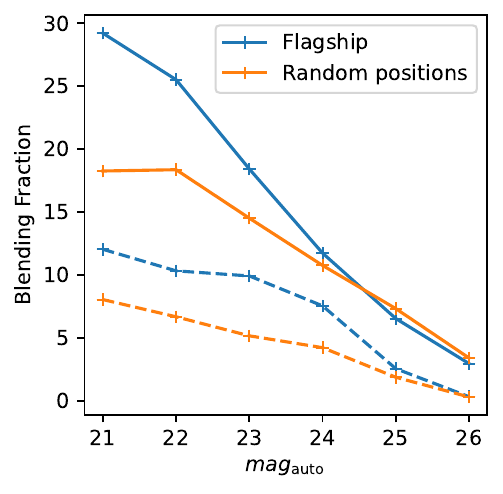}
                \caption{Magnitude-binned blending fraction as seen using the random position simulations and the positions from the Flagship catalogue. Dashed lines show blending as defined by the raising of \SExtractor flag one or two, while solid lines indicate the blending as an overlap of the Kron-ellipses of neighbouring galaxies.}
                \label{fig:Fig15}
            \end{figure}

             We adapted the blending fraction definitions from \citet{blending_synergies}. Thus the blending fraction is defined as the ratio of detected objects that are classified as blends to the total number of detected objects. If one galaxy is blended is determined in two ways. One option is to check if \SExtractor raised one or both of the extraction flags one and two, which indicate an impact from neighbouring objects. The second option is to check if the Kron-ellipses, which \SExtractor determined, overlap. In Fig.~\ref{fig:Fig15} we show both of these definitions binned against $mag_\mathrm{auto}$ for our random position setup and for galaxies placed according to the Flagship catalogue.
            We find that the blending fraction within Flagship is always higher than for random positions for the relevant magnitudes up to 24.5, which shall be used for the cosmological analysis of \Euclid. For the complete sample we find blending fractions of $2.5\%$ ($8.1\%$) for random positions and $4.7\%$ ($10\%$) for Flagship, defined via the \SExtractor flags (Kron-ellipses). Thus by placing galaxies randomly we underestimate the blending fraction by about $2\%$ compared to a realistically clustered case. 

        \subsubsection{Runtime improvements}
            To test the impact that this more realistic clustering has on the runtime improvements, we use our pipeline to study the runtime improvements for our Flagship-based image simulation. We conducted this analysis only for the large shear interval since a similar behaviour is expected for the smaller shear interval as well. In Table~\ref{tab:eff_flagship} the improvements in runtime and area for this new simulation setup are listed. 
            \begin{table}
                \caption{Efficiency comparison for galaxies positioned according to the Flagship mock galaxy catalogue.} 
                \label{tab:eff_flagship} 
                \centering 
                \begin{tabular}{l c c c c c c} 
                \hline\hline 
                Method & \multicolumn{3}{c}{$\mu$-bias} & \multicolumn{3}{c}{$c$-bias} \Tstrut\\
                & RI & $\sigma_\mathrm{RI}$ & AI & RI & $\sigma_\mathrm{RI}$ & AI\\
                \hline %
                \multicolumn{7}{c}{Shear interval $[-0.1, 0.1]$}\Tstrut\Bstrut \\ 
                \hline
                Shape local& 3.5 & 0.3 & 3.5 & 3.0 & 0.3 & 3.0\Tstrut\\
                Both local & 5.8 & 0.6 & 3.5 & 5.3 & 0.4 & 3.2\\
                Shape global & 3.5 & 0.3 & 3.5 & 3.6 & 0.3 & 3.6\\
                Both global & 7.2 & 0.4 & 4.3 & 7.3 & 0.5 & 4.4\\
                RM & 2.5 & 0.2 & 2.5 & 0.1 & 0.01 & 0.1\\
                \hline
                \end{tabular}
                \tablefoot{This table includes galaxies with input magnitudes brighter than 26.5, and the same S/N larger than ten selection. RM stands for the response method, RI for runtime improvement, and AI for area improvement. The uncertainty for the runtime improvement is listed as $\sigma_\mathrm{RI}$.}
            \end{table}
            Comparing this with Table~\ref{tab:eff_rp} we find that all runtimes improvements worsen a little. In absolute terms we find that pixel noise cancellation is more affected than only shape noise cancellation. Also limiting the Flagship catalogue to magnitudes 26.5 and brighter, we find very similar runtime improvements to our random position simulations. Thus we conclude that the reason for the worsening are mainly the additional faint galaxies, which are not detected, but add correlated noise in the background. This mostly affects the pixel noise cancellation, since this additional form of correlated noise by faint galaxies does not get cancelled. In relative terms the response method worsens the most, which we attribute mostly to the absence of the variant of shape noise cancellation we described in Sect.~\ref{sec:4_setup}. The shape noise cancellation we used for the response method on random positions used newly drawn random positions for the rotated galaxies, which made it a softer version of the cancellation compared to local and global cancellation. For the Flagship simulations we omitted this since the shapes and orientations at a certain position in the sky are given by the catalogue. We nevertheless see that our qualitative statements about the most efficient methods still hold. Here we find that both cancellations, especially with the \enquote{global} scene rotation, still provide the best runtime improvement. Thus including clustering and fainter galaxies has only a minor impact on the estimated runtime improvements. Most of the difference that the inclusion of these additional effects causes, is likely absorbed in the simultaneous worsening of the reference efficiency.

\section{Summary and conclusions}\label{sec:7}
    This paper presents two possibilities to improve the efficiency of shear bias calibration simulations, which are indispensable for the scientific analysis of upcoming cosmic shear surveys. One of the methods is pixel noise cancellation, which can be used on top of shape noise cancellation in order to reduce the uncertainties of bias estimation via a fit. This cancellation uses images with an almost exactly inverted noise field in order to cancel noise in the shape estimates caused by the pixel noise realisation. This method is computationally cheap as no further convolution is needed to build the pair image used for cancellation. This advantage can only be fully exploited if the generation of a galaxy image (including the convolution) dominates the runtime compared to the measurement of a galaxy. The advantage therefore depends on the shape measurement methods used for the analysis.
    A completely different approach follows \citetalias{Pujol_2018} in their suggestion to use responses for the bias estimation. We use their ideas to expand their formalism to account for selection bias. We additionally adapt the method to accurately use Poisson noise and make it applicable for larger shear intervals. We compare these two methods with the commonly used fitting of a linear function without any cancellation and the fit using shape noise cancellation. The performance of each method depends on both the chosen shear interval and the type of simulation. Hence we present results in two shear intervals $[-0.1,0.1]$ and $[-0.02, 0.02]$ for both grid-based and random-position simulations. 
    
    In both kinds of simulations, we find that pixel noise cancellation and the response method are very useful for multiplicative bias estimation. Their efficiency is almost the same in the large shear interval, while in small shear intervals the response method has the advantage. The fit naturally depends on the given shear interval, while the response method is largely insensitive to it. Consistently all methods are less effective in the second kind of simulation, where galaxies are placed randomly. We find the largest improvement for the response method on small shear intervals compared to using no cancellation. In this case, the method can improve the runtime by a factor of $145$. In the same shear interval for randomly positioned galaxies, this factor decreases to $70$. Still, this improvement by two orders of magnitude can be beneficial for future bias constraints. In the larger shear interval, we find improvement factors of 13 (grid) and 8 (random positions) for both cancellations and factors of 5 (grid) and 4 (random positions) for the response method. For the additive bias estimation, only the fit method is useful. There the additive bias improvement is the same as the multiplicative bias improvement. The additive bias estimation with the response method is just as good or even worse than using no cancellation. That is because no information for the additive bias can be gained from simulating the same galaxy multiple times with almost the same noise. Implementing pixel noise cancellation for the response method might help improve the capabilities of additive bias estimation. This idea is left for further work. Also an empirical additive bias estimation as suggested in \citet{2021A&A...656A.135H} might be possible. In that case, the response method would not need to be capable of accurately determining the additive bias. 
    
    Our studies of the runtime improvement as a function of magnitude also show that there is no reason not to use pixel noise cancellation on top of shape noise cancellation in any case. It is always at least as efficient as shape noise cancellation, but mostly more efficient. We also find that the advantage of the response method on the small shear interval does not largely depend on the magnitude of the galaxies. This method provides the largest runtime improvement for every magnitude bin on the grid and at random positions. It is solely on the larger shear interval that one must carefully decide if using the response method makes sense. 
    
    Another intriguing effect that we find is the significant dependency of the absolute multiplicative bias estimate on the chosen shear interval. For the particular \KSB shape measurement method used in this study, the multiplicative bias seems to be higher for small shear intervals. This behaviour hints at non-linear bias terms, that are not accounted for by the simple fitting of a linear function. Allowing for an additional quadratic term in the fit, the multiplicative bias changes by an absolute value of $8\times 10^{-3}$. Very recently, this effect of quadratic terms has also been studied by \citet{kitching}. Thus, our current methods of shear bias estimation might not be complete and we may need to quantify biases in the full $g_1$-$g_2$ plane. 
    
    With our findings of potential efficiency improvements for the random position simulations, we highly recommend using pixel noise cancellation. This kind of cancellation is relatively easy to implement and can already reduce the simulation volume needed to reach desired requirements regarding bias uncertainties. Especially in the case of small shear fields, it also makes sense to consider using the response method. Although harder to implement, improving more than two orders of magnitude in runtime can be worth it. 

    When moving to simulations including clustering, we want to highlight the need to quantify the dominant contribution to the runtime for a particular shape measurement method and simulation pipeline in order to decide if it is worth to implement pixel noise cancellation also on top of shape noise cancellation. This is because our simulations with positions drawn from the Flagship catalogue (including clustering) have shown that the area improvements for these simulations are very comparable with and without the additional pixel noise cancellation for our setups with local shape noise cancellation at least. However since \Euclid has a complex PSF that has to be simulated with a higher degree of detail as conducted in our work, it might well be that the PSF generation and convolution becomes a more dominant contribution to the overall runtime, in which case the addition of pixel noise cancellation would provide larger runtime improvements.
    
    In the interest of repeatability and transparency, we make the code publicly available\footnote{\url{https://github.com/HenningJ99/noise_mitigation_code}}. These scripts are not a product of the Euclid Consortium Science Ground Segment. They are created exclusively for the present analysis and made public for reproducibility of the results presented in this paper. We also provide the scripts and data to generate each plot in this paper under the same address.

\begin{acknowledgements}
We thank Andy Taylor, Shun-Sheng Li, and the anonymous referee for their useful comments and discussions about this work. HJ and TS acknowledge support provided by the Austrian Research Promotion Agency (FFG) and the Federal Ministry of the Republic of Austria for Climate Action, Environment, Energy, Mobility, Innovation and Technology (BMK) via the Austrian Space Applications Programme with grant numbers 899537 and 900565, as well as the German
Research Foundation (DFG) under grant 415537506. MT acknowledges support from the German Federal Ministry for Economic Affairs and Climate Action (BMWK) provided by DLR under projects no. 50QE2002 and 50QE2302.
\AckEC 
\end{acknowledgements}

\bibliography{bibliography, euclid}

\onecolumn
\begin{appendix}
  \section{Binned improvements additive bias}\label{sec:app_binned_improvements_c}
\begin{figure*}[!htb]
            \includegraphics{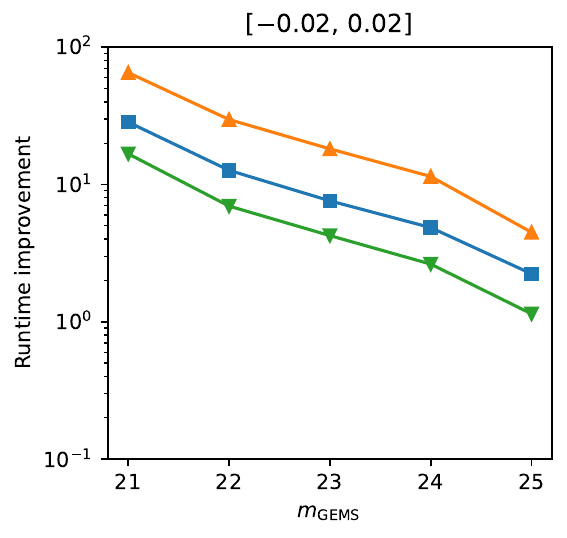}
            \includegraphics{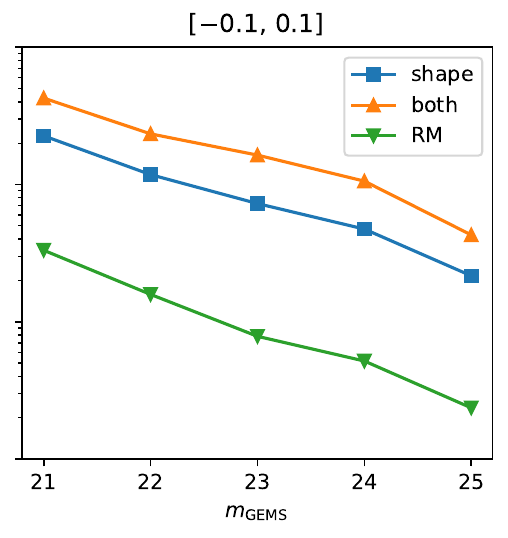}
            \caption{Magnitude-binned runtime improvement of the additive bias for the grid-based simulations. The position of the points marks the center of each bin. The runtime improvement is always compared to the fit method without any cancellation. Error bars are smaller than the symbols and therefore omitted for better visibility.}
            \label{fig:binned_improvement_grid_c}
        \end{figure*}

\begin{figure*}[!htb]
            \includegraphics{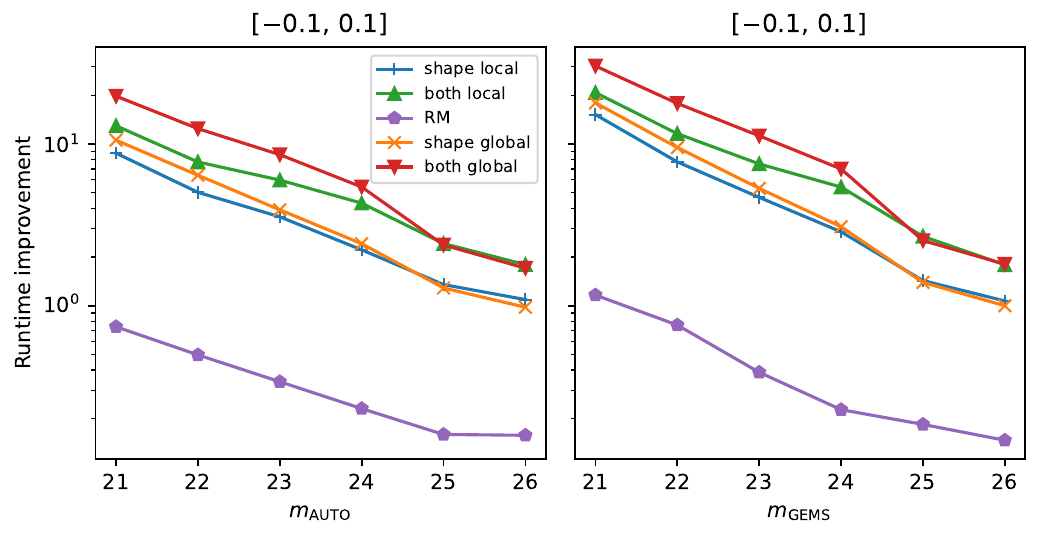}
            \caption{Magnitude-binned runtime improvement of the additive bias for the random position simulations in the large shear interval. In the left panel, the data is binned in $m_\mathrm{AUTO}$ from \SExtractor, while the input magnitude is used for the binning in the right panel to point out the differences. The position of the points marks the center of each bin. The runtime improvement is always compared to the fit method without any cancellation. For this figure, no signal-to-noise cut is applied. Error bars are smaller than the symbols and therefore omitted for better visibility.}
            \label{fig:binned_improvement_rp_c_01}
        \end{figure*}

\begin{figure*}[!htb]
            \includegraphics{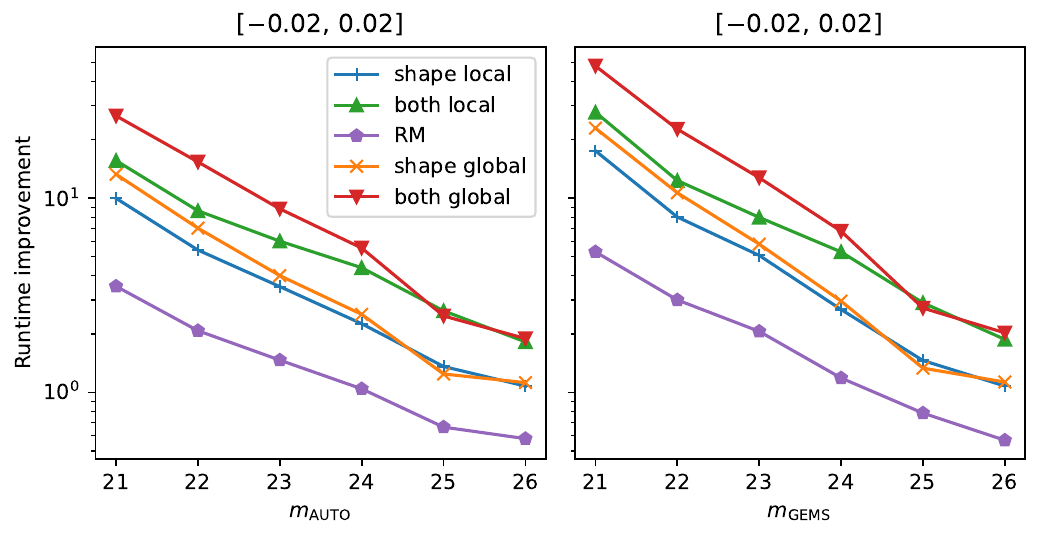}
            \caption{Magnitude-binned runtime improvement of the additive bias for the random position simulations in the small shear interval. In the left panel, the data is binned in $m_\mathrm{AUTO}$ from \SExtractor, while the input magnitude is used for the binning in the right panel to point out the differences. The position of the points marks the center of each bin. The runtime improvement is always compared to the fit method without any cancellation. For this figure, no signal-to-noise cut is applied. Error bars are smaller than the symbols and therefore omitted for better visibility.}
            \label{fig:binned_improvement_rp_c_002}
        \end{figure*}

        The binned improvements can also be studied for the additive bias. In Fig.~\ref{fig:binned_improvement_grid_c} these are shown for the grid simulations and in Fig.~\ref{fig:binned_improvement_rp_c_01} and Fig.~\ref{fig:binned_improvement_rp_c_002} for random positions. 
        \FloatBarrier

    \section{Binned absolute biases}\label{sec:app_absolute_biases} 
\begin{figure*}
            \includegraphics{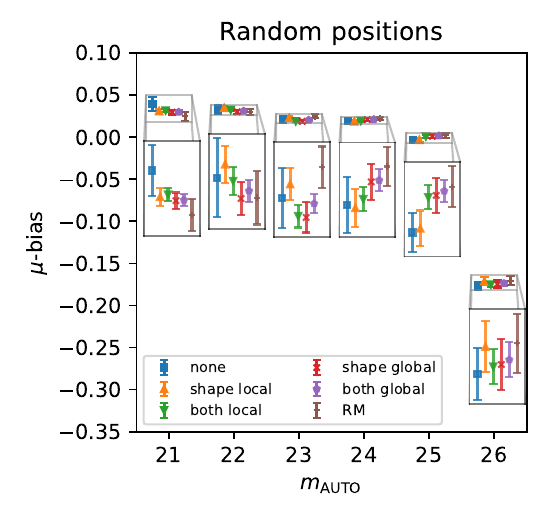}
            \includegraphics{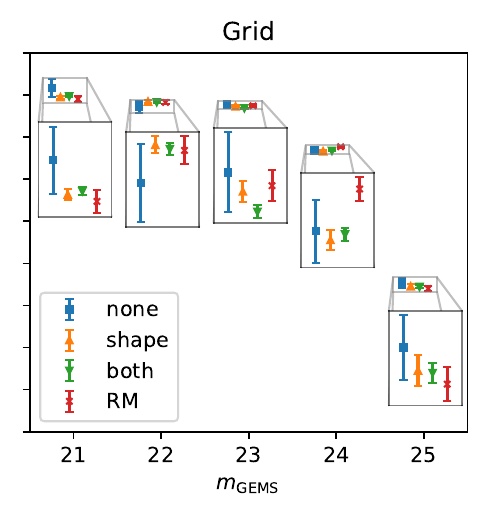}
            \caption{Absolute multiplicative bias comparison in the large shear interval. The general trend of the multiplicative bias at different magnitudes is presented in the main plot. An additional zoomed version of each magnitude bin is shown to better compare the bias estimates since the error bars are too small to be visible in the main plot. For the left panel, no signal-to-noise cut is applied.}
            \label{fig:absolute_biases_m_01}
        \end{figure*}

\begin{figure*}
            \includegraphics{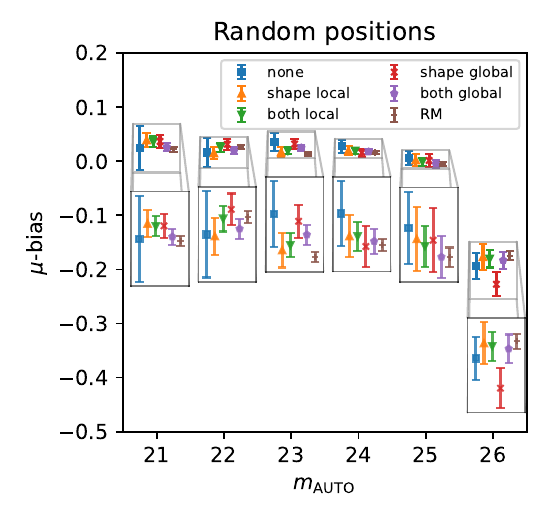}
            \includegraphics{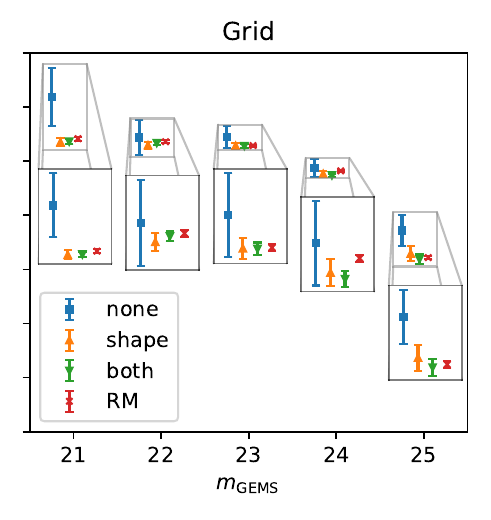}
            \caption{Absolute multiplicative bias comparison in the small shear interval. The general trend of the multiplicative bias at different magnitudes is presented in the main plot. An additional zoomed version of each magnitude bin is shown to better compare the bias estimates since the error bars are too small to be visible in the main plot. For the left panel, no signal-to-noise cut is applied.}
            \label{fig:absolute_biases_m_002}
        \end{figure*}

\begin{figure*}
            \includegraphics{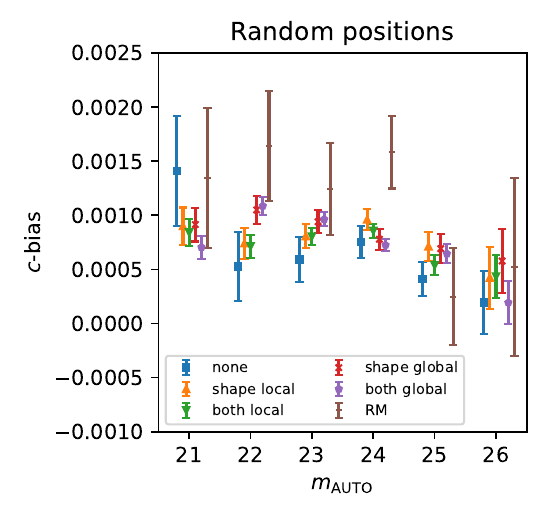}
            \includegraphics{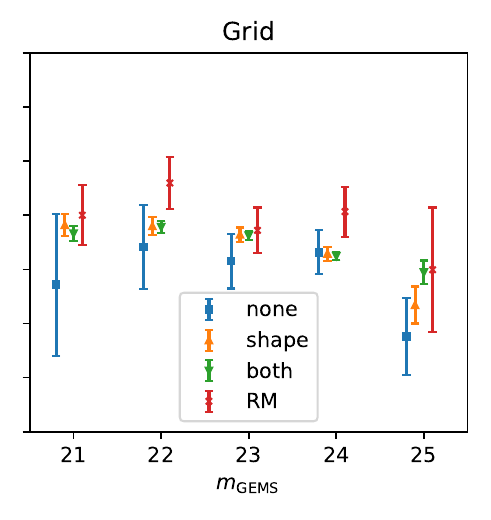}
            \caption{Absolute additive bias comparison in the large shear interval. For the left panel, no signal-to-noise cut is applied.}
            \label{fig:absolute_biases_c_01}
        \end{figure*}

\begin{figure*}
            \includegraphics{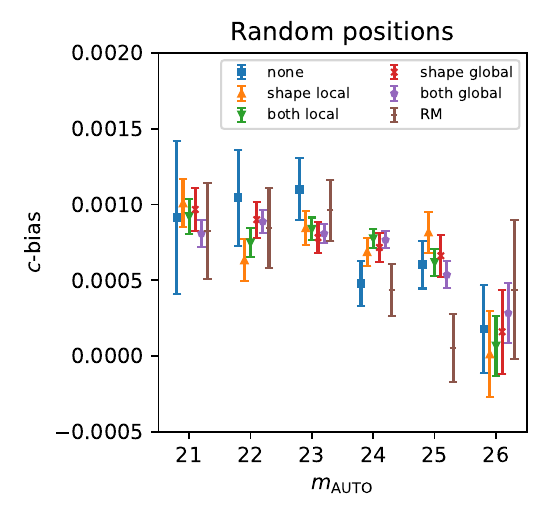}
            \includegraphics{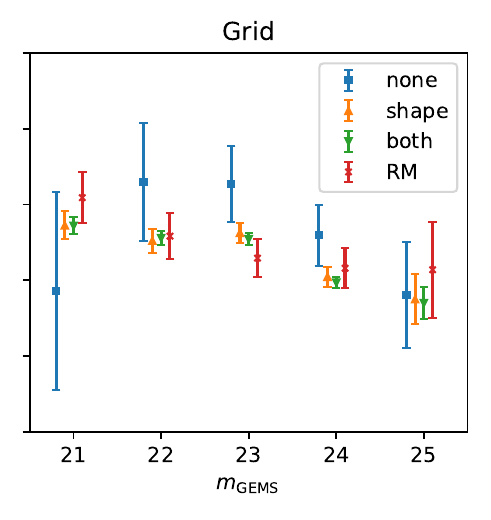}
            \caption{Absolute additive bias comparison in the small shear interval. For the left panel, no signal-to-noise cut is applied.}
            \label{fig:absolute_biases_c_002}
        \end{figure*}

        In order to test if the different methods yield accurate bias estimates, we compare the outcome of the absolute bias measurements for each method in each bin. For the multiplicative bias these comparisons can be seen in Fig.~\ref{fig:absolute_biases_m_01} and Fig.~\ref{fig:absolute_biases_m_002} for the large and the small input shear interval, respectively. For the additive bias they are shown in Fig.~\ref{fig:absolute_biases_c_01} and Fig.~\ref{fig:absolute_biases_c_002}. We observe that the methods are compatible in every magnitude bin, ensuring that every method can be used to determine the biases. The decision of which method to choose only depends on the efficiency of the method.
\end{appendix}

\end{document}